\documentclass{article}

\usepackage[english]{babel}

\usepackage[letterpaper,top=2cm,bottom=2cm,left=3cm,right=3cm,marginparwidth=1.75cm]{geometry}

\usepackage{amsmath,mathtools,amssymb}
\usepackage{graphicx,subfig}
\graphicspath{{Figures/}}
\usepackage{todonotes}
\usepackage[colorlinks=true, allcolors=blue]{hyperref}
\usepackage[dvipsnames]{xcolor}

\usepackage{physics}
\newcommand{\bb}[1]{\mathbf{#1}}
\newcommand{\bs}[1]{\boldsymbol{#1}}

\usepackage[dvipsnames]{xcolor}

\newcommand{\cbox}[1]{#1}

\newcommand{\nablaa}[0]{\nabla_{\!\bb a}}
\newcommand{\nablax}[0]{\nabla_{\!\bb x}}
\newcommand{\Fa}[0]{\bb F\bb a}

\newcommand{\A}{\mathcal{A}}
\newcommand{\B}{\mathcal{B}}
\newcommand{\kH}{k_{\mathrm{H}}}
\newcommand{\Matlab}{Matlab\textsuperscript{\textregistered}}

\usepackage[backend=biber,style=numeric,sorting=none]{biblatex}
\usepackage{siunitx}
\usepackage[normalem]{ulem} 
\usepackage{cases}
\usepackage{enumitem}

\newcommand{\ack}[1]{\paragraph{Acknowledgenmts.} #1}

\title{
    Strongly non-linear continuous model for wave propagation in 2D granular media: derivation and hyperbolicity analysis
}

\author{
Sebastiano Cominelli$^{1,2}$, Régis Cottereau$^{1}$, Sergey Gavrilyuk$^{2}$,
\\ Bruno Lombard$^{1}$, Fabien Petitpas$^{2}$
\\ \small
$^{1}$ Aix Marseille Univ, CNRS, Centrale Med, LMA, Marseille, France
\\ \small
$^{2}$ Aix Marseille Univ, CNRS, Polytech, IUSTI, Marseille, France
}

\begin{document}
\maketitle

\begin{abstract}
    In this work, we develop a nonlinear continuum model for wave propagation in two-dimensional granular materials, building on key ideas from Nesterenko’s work.
    Starting from the Hertzian contact law and neglecting contact friction, we derive an effective continuum description through homogenization, assuming  spherical, monodisperse grains, local periodicity, and the Cauchy--Born hypothesis.
    The resulting strain-gradient model depends on both the first and second gradients of the displacement field. We analyse the hyperbolicity of the leading-order system through its eigenvalues and eigenvectors, establishing conditions for hyperbolicity within a restricted deformation domain. The model is solved using an explicit finite-volume method to investigate gravitational settling and wave propagation. We derive analytical expressions for the longitudinal and transverse wave speeds and highlight their dependence on depth. Finally, the predictions of the continuum model are compared with existing discrete-element simulations.
    This formulation allows several extensions, including more complex packing geometries, contact laws, and large-deformation regimes in which the contact network can evolve.
\end{abstract}

\section{Introduction}

Granular materials are ubiquitous in both natural and engineered environments, with applications ranging from geophysics and civil engineering \cite{forterre2008flows} to pharmaceutical processing \cite{perez2018compaction} and architected materials \cite{leonard2012stress,awasthi2012propagation}. Their mechanical response emerges from the interactions among a large number of discrete particles, giving rise to behaviours that differ fundamentally from those of conventional solids and fluids. Among the many phenomena exhibited by granular media, the study of wave propagation is particularly relevant to geophysical surveying, earthquake engineering, soil characterization \cite{tournat2009nonlinear}, vibration transmission in civil infrastructure \cite{de2017randomly,paolucci2003numerical}, and the mitigation of ground-borne noise and vibrations~\cite{daraio2006energy}.
\\
The discrete nature of granular materials has motivated the widespread use of particle-based models. In particular, the Discrete Element Method (DEM) \cite{cundall1979discrete} explicitly resolves the interactions between individual particles, enabling detailed investigations of wave transmission mechanisms in granular media. Such models have been instrumental in elucidating the role of force chains \cite{owens2011sound,radjai1996force}, contact-network rearrangements \cite{makse2000packing}, depth-dependent wave-speed variations \cite{melin1994wave,jia2020sound,makse2004granular}. As a result, they remain the reference framework for studying wave phenomena at length scales comparable to the grain size.

Despite the level of detail they allow, discrete descriptions are computationally demanding and often provide limited analytical insight into macroscopic wave phenomena. This has motivated the development of continuum theories that seek to represent the collective behaviour of granular assemblies through effective field variables. Continuum descriptions are particularly attractive when the wavelength of interest is significantly larger than the characteristic particle size, allowing homogenization procedures to bridge the microscopic and macroscopic scales.
\\
It is important to distinguish continuum models developed for granular flow from those intended to describe wave propagation. A substantial portion of the granular mechanics literature focuses on the rheology of dense flows \cite{forterre2008flows,lloyd2025formation}, where irreversible rearrangements and friction between the particles dominate the material response. While these models have achieved remarkable success in predicting macroscopic transport and flow phenomena, they are generally not designed to capture the propagation of waves because they rely on mixed hyperbolic--parabolic equations.

Early continuum models for wave propagation focused on one-dimensional granular chains, where the discrete equations of motion were approximated by partial differential equations capable of reproducing dispersion and contact nonlinearity \cite{nesterenko1994solitary,gavrilyuk1996media}. These models provided fundamental insight about the emergence of solitary waves, sonic vacua, and other characteristic features of granular crystals \cite{nesterenko2013dynamics,coste1997solitary,coste1999validity}. Depending on the assumptions adopted during the continuum approximation, different levels of nonlinear behaviour can be captured.
\\
When the particles are assumed to remain in permanent contact under a sufficiently large static precompression, the contact law may be expanded about the equilibrium state, leading to a weakly nonlinear continuum description. Such models accurately describe the propagation of small-amplitude disturbances and are sufficient to reproduce some nonlinear phenomena, including solitary waves.
In contrast, when the wave amplitude becomes comparable to the static confinement, contacts may open and close during the motion, making the full Hertzian interaction essential. This regime is commonly referred to as \emph{strongly nonlinear}. A paradigmatic example is the sonic vacuum, in which an uncompressed granular chain does not support infinitesimal-amplitude waves because the tangent stiffness vanishes in the absence of contact forces. Capturing such phenomena therefore requires retaining the full nonlinear contact mechanics rather than a perturbative approximation about a precompressed state.
\\
The extension of these ideas to two- and three-dimensional granular assemblies is considerably more challenging. Most existing continuum descriptions are derived under the assumptions of weak nonlinearity and small perturbations about a preloaded equilibrium state \cite{muhlhaus1996dispersion,leonard2012stress,awasthi2012propagation,merkel2010elastic,chong2016conical}, or using simplified extensions inspired by one-dimensional settings \cite{favrie2013dynamic}. Such approximations are appropriate for low-amplitude waves, but they become questionable in many situations of practical interest where the wave amplitude may be comparable to the static confinement, contacts may open and close dynamically, and the effective stiffness of the granular assembly evolves together with the wave field. These mechanisms profoundly alter the nature of wave propagation and cannot be captured within weakly nonlinear theories. Despite the extensive literature on strongly nonlinear dynamics in one-dimensional chains, a multidimensional continuum framework capable of describing these effects is still lacking.

At the same time, there is growing consciousness that granular materials cannot be fully described within the framework of classical Cauchy continua \cite{suiker2001micro,gavrilyuk1996media}. Higher-order theories, often referred to as second-gradient or generalised continua, are therefore attracting increasing attention. Such models naturally incorporate characteristic microstructural length scales and provide a continuum description of dispersive phenomena arising from the discrete nature of granular assemblies.
A variety of generalised continuum theories have been proposed, including Cosserat, micromorphic \cite{muhlhaus1996dispersion}, strain-gradient \cite{liao2000pressure,suiker2001micro,placidi2021micromechanics}. These higher-order continuum models are derived from micromechanical considerations, but are based on either linear interaction laws, linear (or weakly nonlinear) kinematics, or both. 
For example, the continuous equations have been derived by linearising the equation of motion in a particular reference configuration \cite{liao2000pressure} or by assuming quadratic energies at the continuous level \cite{muhlhaus1996dispersion,yu2022micromorphic}, or both \cite{placidi2021micromechanics}.
A continuum theory capable of simultaneously incorporating non-local effects, multidimensional microstructural features, and strong nonlinearity is missing.
\\
Existing continuum models are typically derived for a specific reference configuration, most commonly assuming particles arranged on a regular hexagonal \cite{gilles2003low,hua2019wave,merkel2010elastic,chong2016conical} or square lattice \cite{awasthi2012propagation,leonard2012stress,leonard2013directional}. Although most of these regular packings can be mapped onto one another through finite deformations, such transformations profoundly modify the contact network and, consequently, the propagation of elastic waves. This sensitivity has been clearly demonstrated in the experimental investigations of Zhu \emph{et al.}~\cite{zhu1996effect}, who showed that changes in particle orientation and contact topology significantly alter the dynamic response of granular assemblies. To the best of our knowledge, a continuum model capable of capturing this geometric, or kinematic, nonlinearity has not yet been developed.

In this work, we develop a continuum framework for wave propagation in two-dimensional granular materials that is applicable in the strongly nonlinear regime. The proposed formulation retains both the nonlinear Hertzian contact and the nonlinear grain kinematics responsible for the evolution of the contact network. Assuming that the characteristic wavelength is large compared with the particle size, we adopt the Cauchy--Born hypothesis \cite{ericksen2008cauchy}, whereby the granular assembly deforms locally as a periodic lattice. Starting from the Hertzian interaction energy, we derive an effective constitutive law and show that the granular medium can be described as a nonlinear hyperelastic continuum.
\\
The homogenization procedure is carried out to different orders in the particle size. At leading order, the resulting continuum is governed by a non-convex strain-energy density, yielding a system of conservation laws that is only conditionally hyperbolic. The loss of hyperbolicity depends on the deformation state and originates from the transverse motion of particles relative to the contact directions, a mechanism that is overlooked in one-dimensional granular chains. Retaining higher-order terms enriches the model with dispersive effects associated with the underlying microstructure. These non-local contributions introduce an intrinsic length scale and regularize the dynamics, in a manner analogous to the role played by strain-gradient terms in the classical Ericksen bar model \cite{ericksen1992introduction}.

The remainder of the paper is organised as follows. Section~\ref{sec:homogenization} presents a homogenization procedure based on the Hertzian contact energy. The resulting continuum description reproduces the discrete dynamics up to second order in the particle size and naturally takes the form of a second-gradient continuum. The derivation extends Nesterenko's classical one-dimensional framework to two and three dimensions, while recovering the original model as a particular case of one-dimensional motion. In contrast to previous approaches based on the equations of motion, the present formulation is derived directly from the interaction energy.
Section~\ref{sec:hyperbolicity} is devoted to the mathematical analysis of the governing equations. Particular attention is paid to the leading-order system, which is shown to be conditionally hyperbolic, with the domain of hyperbolicity depending on the contact configuration and state of deformation.
In Section~\ref{sec:numerics}, the leading-order equations are discretised using a finite-volume scheme. Numerical simulations were performed on a granular assembly undergoing gravitational settling. This resulted in a non-uniform stress distribution. Wave propagation is then investigated in a configuration inspired by the numerical experiments of Melin~\cite{melin1994wave}.
Finally, Section~\ref{sec:conclusions} summarises the main results and discusses perspectives for future developments.

\section{Continuous Lagrangian of a 2D stack of grains}
\label{sec:homogenization}

\begin{figure}
    \centering
    \subfloat{\includegraphics[scale=.35,trim=50 60 50 60,clip]{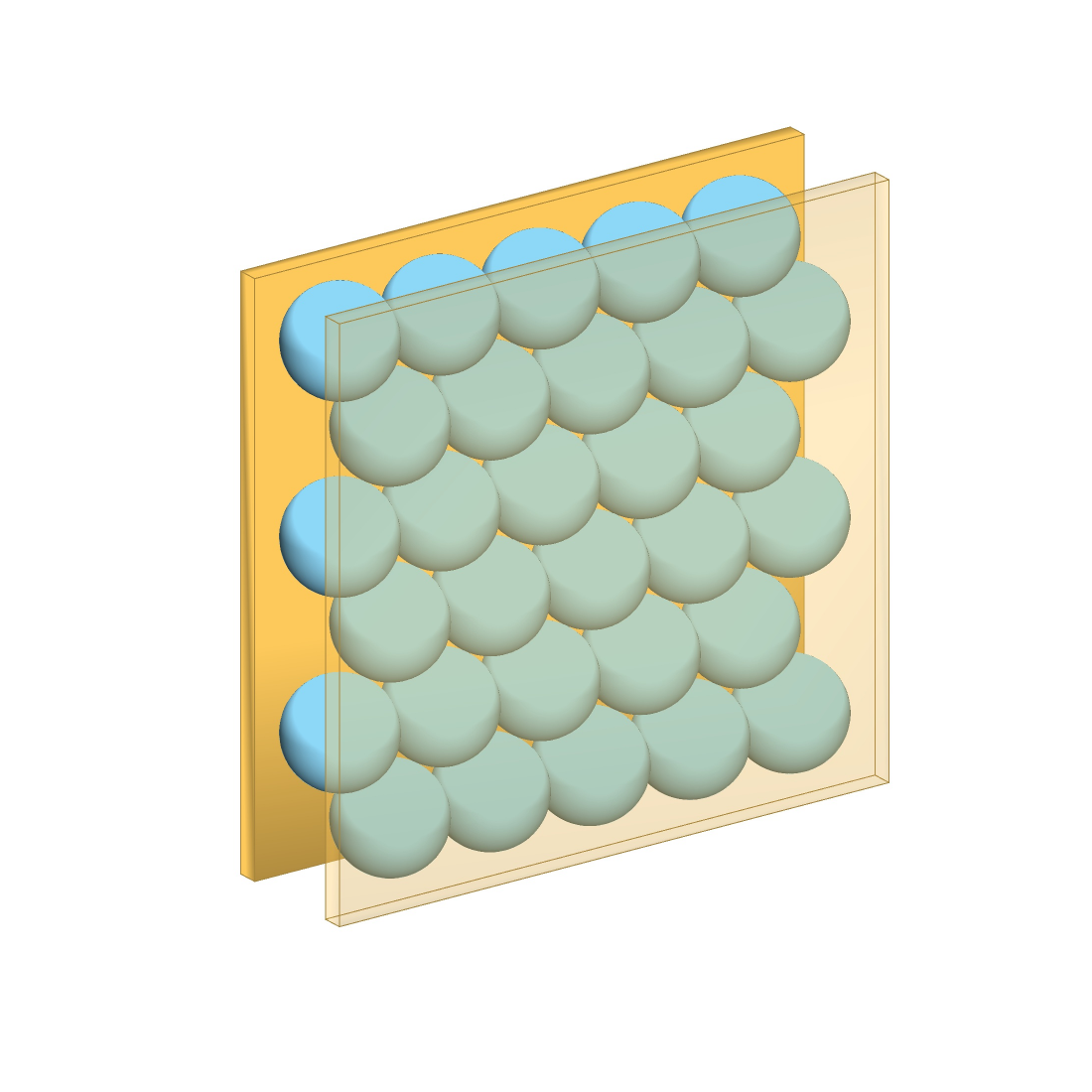}\label{fig:hex_pack_sph}}\;
    \subfloat{\includegraphics[scale=.35,trim=50 60 50 60,clip]{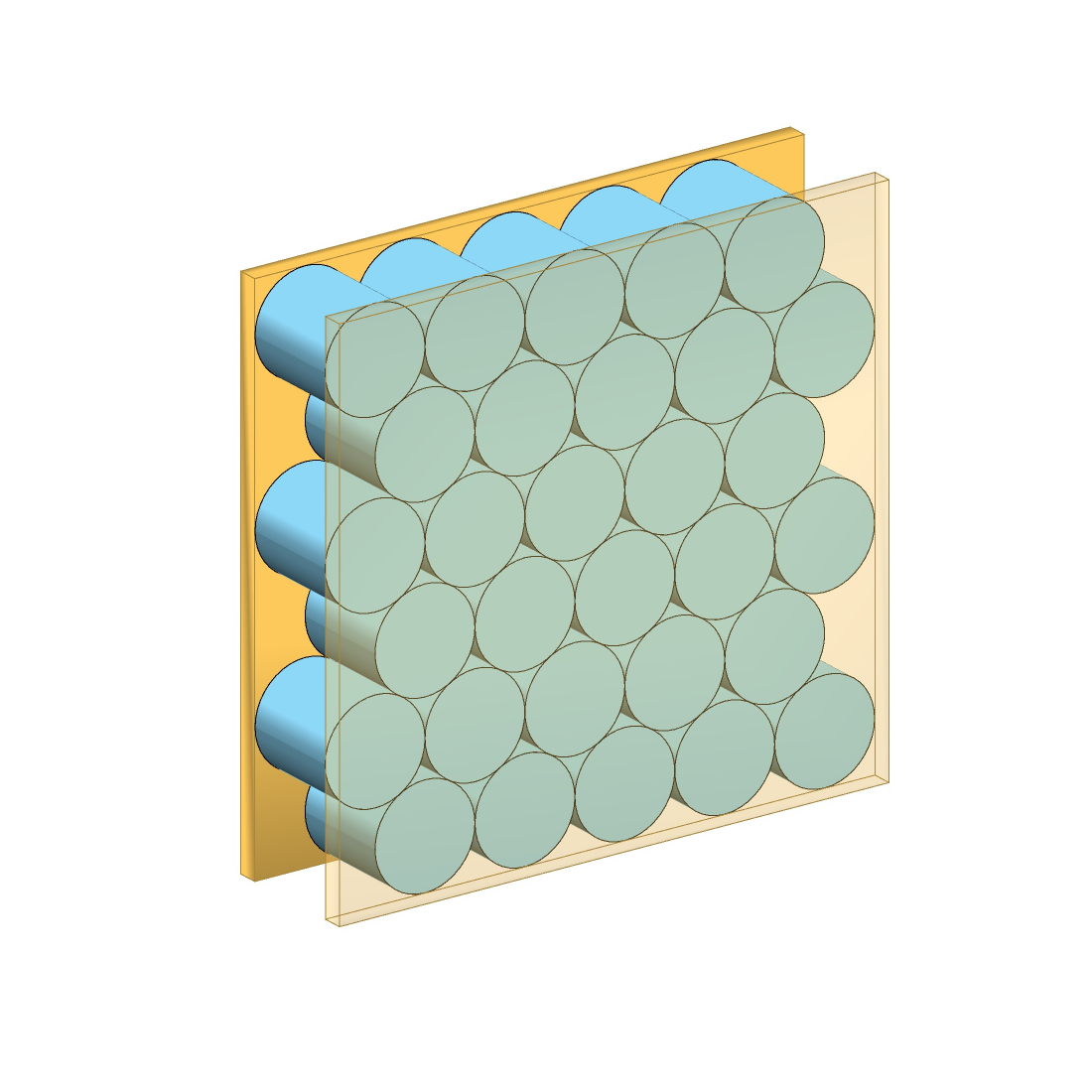}
    \label{fig:hex_pack_cyl}}
    \caption{Honeycomb packings of spheres and cylinders confined by two parallel plates.}
    \label{fig:hex_pack}
\end{figure}

\begin{figure}
    \centering
    \subfloat[]{\includegraphics[scale=.6,trim=-20 5 0 0]{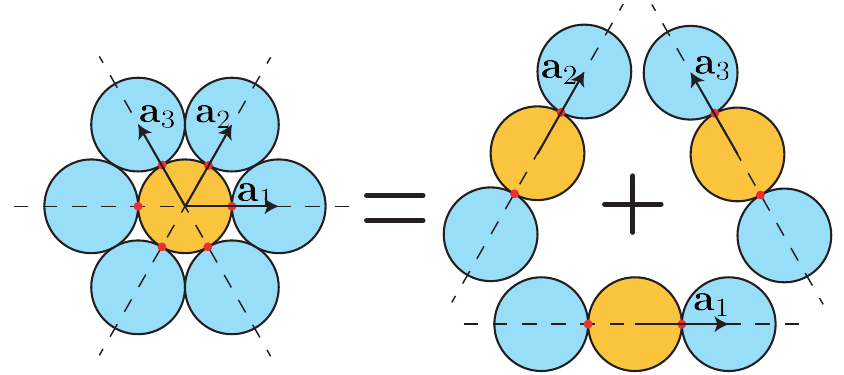}\label{fig:ref conf}}\;
    \subfloat[]{\includegraphics[scale=.6,trim=-10 5 0 0]{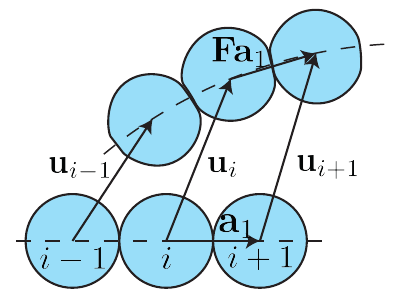}\label{fig:schematic}}
    \caption{(a) One grain's contacts in the reference configuration are classified along three directions; (b) schematic of the motion of a chain of grains in two dimensions.}
\end{figure}

Let us consider a periodic arrangement of particles moving in a plane, with their centres arranged in a hexagonal packing, as illustrated in Figure~\ref{fig:hex_pack} for spherical and cylindrical particles. Such configurations can be realised, for instance, by confining the beads between two rigid parallel planes. We focus on spherical particles, for which the contact forces follow a nonlinear Hertz law, in contrast to cylindrical particles with parallel, aligned axes, for which the contact response is linear. This configuration is similar to that studied by Melin~\cite{melin1994wave} using a discrete model.
\\
We identify as the \emph{reference configuration} the one schematised in Figure~\ref{fig:ref conf}, where each sphere is barely in contact with its six nearest neighbours, and we recognise three directions of contact along the vectors $\bb a_1$, $\bb a_2$, and $\bb a_3$ that connect the centres of two neighbours.
In search for a continuous model, we describe the motion of each particle through effective kinetic $\mathcal K$ and potential $\mathcal V$ energies, and  define a density Lagrangian in the form $\mathcal L = \mathcal K - \mathcal V$.
The equations of motion for a continuous material will be derived using the principle of least action.
\\
For moderate forces, the particles are mainly deformed around the contact points, and elastic deformation in the rest of the particle can be neglected to a good approximation.
Remarkably, this regime allows us to adopt the Hertz's law.
We also assume that the contact in one point does not influence any other contact in the spheres.
In practice, this mutual interaction is in the order below of \SI{1}{\percent}, as one can prove following \cite{hiramatsu1965determination,liu2010elastic}.
This observation allows us to consider the potential energy stored in the three directions of contact independently.
For example, the potential energy stored in the six contacts of the central sphere in Figure~\ref{fig:ref conf} can be seen as the sum of the potential energies stored in the three highlighted pairs of contacts. Finally, no friction between the particles is modelled.

The method we follow to derive a continuous model is similar to the one firstly adopted by Nesterenko~\cite{nesterenko2013dynamics} to model the motion of one-dimensional chain of grains, but it differs in the order of derivation.
In such a classical approach, one starts from the equilibrium of contact and the inertia forces acting on a grain. This equation, written for one representative grain in discrete form, is homogenised under the long-wavelength assumption, obtaining a continuous momentum equation. Finally, a continuous energy formulation is derived by integration, obtaining the Lagrangian density function of the system.
Conversely, we start from a discrete form of the energy and make it continuous through homogenization. Then, we derive the equation of motion at the continuous level.
The reader is referred to Triantafyllidis \emph{et al.}~\cite{triantafyllidis1993higher} for a general discussion on this approach in one-dimensional spring--mass chains.
\\
Starting from the energy is required in two or three dimensions because one has to model two types of nonlinearities: the Hertz's law and the geometric motion. Since homogenization is performed by carrying Taylor expansions, by starting from the vectorial force equilibrium, one is forced to expand in polynomial form both nonlinearities and choose a truncation. Conversely, by starting from the energy, which is a scalar function, one can expand only the Hertz's law and truncate the series at the desired order; the nonlinear equations of motion are derived after the homogenization, without any approximation on the kinetics of the motion.
\\
In this section, all the computations are done assuming a two-dimensional space, but the generalisation to three dimensions is straightforward.

\subsection{Effective kinetic energy}
The kinetic energy of one grain is assumed to depend only on the velocity of its centre of mass. No deformations inside the grain or rotations are accounted in this model. Indeed, the rotation of a frictionless spherical particle cannot be influenced by any contact.
We define the reference effective density $\rho_0$ computed as the spatial average of the granular assembly in the reference configuration, where the voids between two grains are considered massless and the grains have a constant mass $\rho_\mathrm{p}$.
In one dimension, when only one chain of spheres is considered, it is $\rho_{0,\mathrm{lin}}=\frac\pi6\rho_\mathrm{p}\approx0.524\,\rho_\mathrm{p}$; in two dimensions, by considering the hexagonal pattern described above, we have $\rho_{0,\mathrm{hex}}=\frac{\sqrt3\pi}{9} \rho_\mathrm{p}\approx0.605\,\rho_\mathrm{p}$.
Note that these values refer to packing of \emph{spheres} and are different from those for packing of \emph{disks}. The latter would be obtained by considering only the two-dimensional filling fractions: $\frac\pi4\approx0.785$ for the one-dimensional chain and $\frac\pi{2\sqrt3}\approx0.907$ for the hexagonal pattern.
Thus, the kinetic energy of grains in Lagrangian coordinates is
\begin{equation}
    \mathcal K=\frac12 \rho_0|\bb v|^2,
\end{equation}
where $\bb v$ is the velocity field.

\subsection{Potential energy of a chain of grains in two dimensions}
When two particles come into contact, a normal force develops due to their elastic deformation. We model such an interaction using the Hertzian contact theory. For two identical spheres, the normal force $F_\mathrm{H}$ is related to the amount of contact $\delta$ through
\begin{equation}\label{eq:Hertz}
    F_\mathrm{H}(\delta) = \frac{\sqrt2E_\mathrm{p}}{3(1-\nu_\mathrm{p}^2)} \sqrt R \;(\,\delta\,)_+^{3/2},
\end{equation}
where $E_\mathrm{p}$ and $\nu_\mathrm{p}$ are the Young's modulus and the Poisson's ratio of the material composing the particles, respectively. $R$ is the radius of the spheres.
The notation $(\,\cdot\,)_+$ indicates zero when the argument is negative, and the value of the argument otherwise.
Note that, despite its piecewise definition, the function $F_\mathrm{H}(\delta)$ is continuously differentiable.
In the present work, only normal contact forces are considered as we assume that they dominate the wave dynamics, being responsible for the strongly nonlinear phenomena of primary interest, including solitary-wave propagation. Other interaction mechanisms, such as friction, tangential contact forces, and moment transfer, have been incorporated into more elaborate discrete models, often through linear constitutive laws (see, e.g., \cite{suiker2001micro}). Moreover, for circular particles, normal contact forces do not generate contact moments.
\\
The energy stored in the contact is computed as
\begin{equation}
    E = \int_0^\delta F(\tilde\delta)\,\mathrm d\tilde\delta=\frac{\sqrt2E_\mathrm{p}}{3(1-\nu_\mathrm{p}^2)} \sqrt R \,\frac{\big(\delta\big)_+^{n+1}}{n+1},
\end{equation}
where $n=3/2$, corresponding to the Hertzian contact law for spheres, and we retain the exponent $n$ throughout the derivation to preserve generality. In particular, $n=1$ recovers a linear contact law, representative of cylindrical contacts, and $n=2$ can account for sphere irregularities or prismatic grains \cite{coste1999validity,goddard1990nonlinear}.
Thus, the energy specific to the mass of the two particles involved in the contact is
\begin{equation}
    \mathcal E = \frac{E}{2\,\frac43\pi R^3\rho_\mathrm{p}} =
    \frac{\kH}{2(n+1)}\,\big(\delta\big)_+^{n+1},
    \quad\text{with }\;\; \kH=\frac{\sqrt2 E_\mathrm{p}}{4\pi\rho_\mathrm{p} (1-\nu_\mathrm{p}^2) \, R^{5/2}}.
\end{equation}
This is the energy stored in one contact, when the discrete system is considered. Below, we derive the continuous energy $e$ of an effective medium through homogenization, and obtain $\mathcal V=\rho_0 e$.

Let us consider Figure~\ref{fig:schematic}, where the motion of a chain of grains is represented between the reference and the actual configurations.
The contact energy $e_i$ accumulated in the particle $i$ is the sum of the energies due to the contacts with particles $i-1$ and $i+1$:
\begin{align}\label{eq:NRJ-Hertz}
    e_i &= \frac{\kH}{2(n+1)}
    \Big[
    \Big(|\bb a| -|\bb a + \bb u_i-\bb u_{i-1}| \Big)^{n+1}_+
    +
    \Big( |\bb a| -|\bb a + \bb u_{i+1}-\bb u_i|\Big)^{n+1}_+
    \Big],
\end{align}
where $\bb u_i$ is the displacement of the particle $i$. The vector $\bb a$ is the lattice vector shown in Figure~\ref{fig:schematic} where we have dropped the subscript 1 to simplify the notation throughout this section.
\\
The following derivation does not require any bound on the magnitude of the displacement $\bb u_i$. However, by writing the potential energy~\eqref{eq:NRJ-Hertz} we assume  that interactions are limited to pairs of particles that are already in contact in the reference configuration.
In principle, the energy should account for all possible particle interactions, since sufficiently large deformations may bring initially distant particles into contact. In other words, the adjacency graph of the contact network is assumed to remain unchanged throughout the deformation.
This restriction has no counterpart in one-dimensional granular chains, where the ordering of the interacting neighbours is fixed throughout the motion.
This assumption is expected to hold for moderate deformations, making it well suited to the study of wave propagation. In the numerical simulations, its validity will be verified \emph{a posteriori}, monitoring that no contact is lost in the sense described in Section.~\ref{sec:3 contacts}.

\textbf{Long-wavelength assumption.}
We adopt the Cauchy--Born hypothesis, in which the deformation of the granular assembly is assumed to be locally homogeneous at the particle scale. Under this assumption, the motion of each particle is entirely determined by the macroscopic deformation gradient, allowing the discrete system to be replaced by an equivalent continuum. Although originally introduced for crystalline solids, the Cauchy--Born rule has become a standard homogenization tool in lattice mechanics. A comprehensive review of its mathematical foundations and range of validity is provided by Ericksen \cite{ericksen2008cauchy}.
In the granular mechanics literature, the same assumption is also referred to as the \emph{kinematic hypothesis}~\cite{suiker2001micro} or \emph{Piola's ansatz}~\cite{placidi2021micromechanics}. It is generally recognized that this approximation tends to overestimate the stiffness of the discrete assembly, in contrast to the complementary \emph{static hypothesis}, which typically provides a lower bound. The reader is referred to Liao \emph{et al.}~\cite{liao2000pressure} for a detailed discussion and more references.

Let $\bb u (\bb X,t)$ be a smooth field that interpolates the displacement of every particle when evaluated in the particle position $\bb X_i$, i.e.\ $\bb u_{i}(t)= \bb u(\bb X_{i},t)$.
When the motion is described by a plane wave, i.e.\ $\bb u(\bb X,t)=\tilde{\bb u}\,e^{i(\omega t-\bb k\cdot x)}$, the differential operator $\nabla$ (computed with respect to the Lagrangian coordinates) is of the order of the wavenumber $\bb k$. Thus
\begin{equation}
    \nabla\bb u\cdot\bb a =\nablaa\bb u \sim (\bb k\cdot\bb a) \bb u,
\end{equation}
where $\nablaa\coloneqq\bb a\cdot\nabla$ is the directional derivative in Lagrangian coordinates along the lattice vector $\bb a$. 
We choose $\varepsilon \coloneqq|\bb k| |\bb a|$ and carry out the long-wavelength homogenization by considering $\varepsilon\ll1$.
If $\bb k$ is parallel to $\bb a$, this choice means that the wavelength is much larger than the size of the grains;
otherwise, $\bb k\cdot\bb a<\varepsilon$ and $\varepsilon$ overestimates $\bb k\cdot\bb a$; in any case, we can highlight different orders of $\varepsilon$ by performing a Taylor's expansion and recognising that $\nablaa= o(\varepsilon^0)$,  $\nablaa^2= o(\varepsilon)$, etc.
In the following, we do not highlight the small parameter $\varepsilon$, but we keep it inside the derivative operators.

We expand around $\bb X_i$ to obtain the displacement of the neighbouring particles labelled $i\pm1$:
\begin{equation}
    \bb u_{i\pm1}
    \approx
    \bb u(\bb X_i)
    \pm \nablaa \bb u(\bb X_i)
    + \tfrac12 \nablaa^2 \bb u(\bb X_i)
    \pm \tfrac16 \nablaa^3 \bb u(\bb X_i)
    + o(\varepsilon^3),
\end{equation}
where $\nablaa^k$ is the directional derivative along $\bb a$ of order $k$.
Substituting into Eq.~\eqref{eq:NRJ-Hertz}, we obtain
\begin{equation}
    e = \tfrac{\kH}{2(n+1)}
    \Big[
    \Big(
    |\bb a|
    - |\bb a + \nablaa \bb u
    - \tfrac12 \nablaa^2\bb u
    + \tfrac16 \nablaa^3\bb u |
    \Big)^{n+1}_+
    \hspace{-5pt}+    
    \Big(
    |\bb a|
    - |\bb a + \nablaa\bb u
    + \tfrac12 \nablaa^2\bb u
    + \tfrac16 \nablaa^3\bb u |
    \Big)^{n+1}_+
    \Big],
\end{equation}
where the argument $\bb X_i$ has been omitted to ease the notation.
By definition, $\bb u = \bb x - \bb X$, then
\begin{align}
    \nabla \bb u &= \bb F - \bb I, &
    \nablaa \bb u &= \Fa - \bb a, &
    \nablaa^2 \bb u &= \nablaa \Fa, &
    \nablaa^3 \bb u &= \nablaa^2 \Fa;
\end{align}
thus,
\begin{equation}
    e = \frac{\kH}{2(n+1)}
    \Big[
    \Big(|\bb a| - |\Fa
    - \tfrac12 \nablaa \Fa
    + \tfrac16 \nablaa^2\Fa+\dots|\Big)^{n+1}_+
    + \Big(\cdots\Big)^{n+1}_+
    \Big].
\end{equation}
To correctly truncate the expansion, we evaluate the orders of magnitude of the following terms
\begin{equation}
    \Fa = \bb a + \nablaa \bb u
    = O(|\bb a|) + o(\varepsilon^0),
    \quad
    \nablaa \bb F \bb a = o(\varepsilon),
    \quad
    \nablaa^2 \bb F \bb a = o(\varepsilon^2).
\end{equation}
We define
\begin{equation}
    \bs\psi^\pm = \pm\frac12 \nablaa \Fa + \frac16 \nablaa^2\Fa \pm \frac1{24}\nablaa^3\Fa + o(\varepsilon^4).
\end{equation}
Then,
\begin{equation}\label{eq:energy}
    e =
    \frac{\kH}{2(n+1)}
    \Big[
    \Big(|\bb a|-|\Fa+\bs\psi^-|\Big)_+^{n+1}
    +
    \Big(|\bb a|-|\Fa+\bs\psi^+|\Big)_+^{n+1}
    \Big].
\end{equation}
\textbf{Taylor expansion of the energy.}
We define
\begin{equation}
    g(\bs\psi)\coloneqq\frac{\kH}{n+1}\Big(|\bb a|-|\Fa+\bs\psi|\Big)^{n+1}_+
\end{equation}
and perform the Taylor expansion around $\bs\psi=0$:
\begin{equation}
    g(\bs\psi)
    =
    g(\bs0)
    +
    \frac{\partial g(\bs0)}{\partial\bs\psi}\cdot \bs\psi
    +
    \frac12
    \frac{\partial^2 g(\bs0)}{\partial\bs\psi^2}:\bs\psi\otimes\bs\psi
    +\dots
\end{equation}
Remarkably, by expanding the energy we approximate only the Hertzian nonlinearity, while keeping untouched any geometric nonlinearity related to the kinematics.
We note that
\begin{equation}
    \frac{\partial g(\bs0)}{\partial\bs\psi}
    =
    \frac{\partial h(\Fa)}{\partial \Fa},
    \quad \text{with}\quad
    h(\Fa)\coloneqq\frac{\kH}{n+1}\Big(|\bb a|-|\Fa|\Big)^{n+1}_+.
\end{equation}
For the sake of brevity, we define $\bb h \coloneqq\frac{\partial h}{\partial \Fa}$ and $\bb H \coloneqq\frac{\partial^2 h}{\partial\Fa^2}$:
\begin{subequations}
\begin{align}
    &\bb h =
    -\kH\frac{(|\bb a|-|\Fa|)^{n}_+}{|\Fa|}\Fa,
    \\
    &\bb H  =
    \kH\Big(n\frac{(|\bb a|-|\Fa|)^{n-1}_+}{|\Fa|^2} \Fa\otimes\Fa
    -\frac{(|\bb a|-|\Fa|)^n_+}{|\Fa|} \bb I
    +\frac{(|\bb a|-|\Fa|)^n_+}{|\Fa|^3}  \Fa\otimes\Fa\Big).
\end{align}
\end{subequations}
Nothing can be said \emph{a priori} on the order of magnitude of these derivatives, thus we stay conservative and consider them all of $O(\varepsilon^0)$.
We expand the energy \eqref{eq:energy} as
\begin{align}
    e =\frac12\big[g(\bs\psi^-)+g(\bs\psi^+)\big]
    &=\frac12\big[
    h
    +\bb h\cdot\bs\psi^-
    +\frac12 \bb H:\bs\psi^-\otimes\bs\psi^- +
    h
    +\bb h\cdot\bs\psi^+
    +\frac12 \bb H:\bs\psi^+\otimes\bs\psi^+
    \big]
    \notag \\
    &=
    h
    +\frac12\bb h\cdot(\bs\psi^-+\bs\psi^+)
    +\frac14 \bb H:(\bs\psi^-\otimes\bs\psi^- +\bs\psi^+\otimes\bs\psi^+),
\end{align}
and compute
\begin{align}
    \bs\psi^-+\bs\psi^+ &=
    \frac13\nablaa^2\Fa + o(\varepsilon^4)
    \notag\\
    \bs\psi^-\otimes\bs\psi^-+\bs\psi^+\otimes\bs\psi^+
    &=
    \frac14\nablaa\Fa\otimes\nablaa\Fa 
    -\frac1{12}\nablaa^2\Fa\otimes\nablaa \Fa
    -\frac1{12}\nablaa\Fa\otimes\nablaa^2\Fa
    + o(\varepsilon^5)
    \notag\\
    &
    \quad+\frac14\nablaa\Fa\otimes\nablaa\Fa 
    +\frac1{12}\nablaa^2\Fa\otimes\nablaa \Fa
    +\frac1{12}\nablaa\Fa\otimes\nablaa^2\Fa
    + o(\varepsilon^5)
    \notag\\
    &=\frac12\nablaa\Fa\otimes\nablaa\Fa
    + o(\varepsilon^5)
\end{align}
Thus, the potential energy is
\begin{equation}
    e = 
    h
    +\frac16 \bb h\cdot\nablaa^2\Fa
    +\frac18 \bb H:(\nablaa\Fa\otimes\nablaa\Fa)
    +o(\varepsilon^4).
\end{equation}
This energy depends on $\bb F$, $\nabla\bb F$ and $\nabla\nabla\bb F$ that are related to the first, the second, and the third gradients of $\bb u$, respectively. Thus, the effective material may be interpreted as a third-gradient material. However, we can get rid of the third gradient performing integration by parts:
\begin{equation}
    \int_t\int_\Omega \bb h\cdot\nablaa^2\Fa\,\mathrm d\Omega\,\mathrm dt
    =
    \int_t\int_{\partial\Omega} (\bb h\cdot\nablaa\Fa)\,\bb a\cdot\bb n\,\mathrm dS\,\mathrm dt
    - \int_t\int_\Omega \bb H:(\nablaa\Fa\otimes\nablaa\Fa)\,\mathrm d\Omega\,\mathrm dt
\end{equation}
where the Gauss--Ostragradsky's theorem has been applied with respect to the directional derivative $\nablaa$, and the chain rule $\nablaa \bb h=\bb H\cdot\nablaa\Fa$ has been used.
The boundary integral vanishes because the deformation is assumed to have compact support, and we obtain the energy
\begin{equation}
    e = 
    h
    -\tfrac1{24}\bb H:\nablaa\Fa\otimes\nablaa\Fa
    +o(\varepsilon^4).
\end{equation}

\subsection{Lagrangian for a 2D chain of grains using Boussinesq trick}

To derive the equation of motion, we consider the kinetic energy and define the Lagrangian
\begin{equation}\label{eq:Lagr}
    \mathcal L \coloneqq
    \frac12 \rho_0 |\bb v|^2 - \rho_0 e(\bb F,\nabla\bb F) = \frac12\rho_0 |\bb v|^2 - \rho_0\Big(h-\frac{1}{24}\bb H:\nablaa\Fa\otimes\nablaa\Fa\Big).
\end{equation}
Thanks to the Cauchy--Born hypothesis, the homogenization we have performed for the particle in position $\bb X_i$ is the same for all other particles and the Lagrangian~\eqref{eq:Lagr} describes the motion of the whole volume under consideration.
\\
The sign of matrix $\bb H$ is indefinite. Indeed, given a vector $\bb w\in\mathbb R^2$, $\bb w\cdot\bb H\bb w \ge0$  if $\bb w$ is parallel to $\Fa$, while $\bb w\cdot\bb H\bb w\le0$ if $\bb w$ is orthogonal to $\Fa$.
Thus, the potential energy derived above contains a higher-order term that is potentially negative, rendering the Lagrangian ill-posed at sufficiently high spatial frequencies. To circumvent this difficulty, we adopt the classical Boussinesq trick (see, e.g., \cite{allaire22AA}). The underlying idea is to exploit the leading-order equations of motion to replace higher-order spatial derivatives with equivalent higher-order time derivatives. Since the two formulations differ only by terms of higher asymptotic order, they remain equivalent within the accuracy of the homogenization, while the resulting model becomes well posed and preserves a correct long-wavelength formulation.
We consider the leading-order approximation of the energy:
\begin{equation}\label{eq:pot eng}
    e = h(\Fa) + o(\varepsilon^3).
\end{equation}
By applying the least action principle with potential energy \eqref{eq:pot eng}, we obtain the momentum equilibrium that, written in index notation for the sake of clarity, reads as:
\begin{equation}
    \ddot u_i =\frac{\partial}{\partial X_j}\frac{\partial e}{\partial F_{ij}} =
    \frac{\partial}{\partial X_j}\Big(\frac{\partial e}{\partial F_{kl} a_l}\frac{\partial F_{km}a_m}{\partial F_{ij}}\Big) =
    \frac{\partial}{\partial X_j} \Big(\frac{\partial e}{\partial F_{il}a_{l}}a_j\Big)=
    a_j\frac{\partial}{\partial X_j}\frac{\partial (h+o(\varepsilon^3))}{\partial F_{il}a_l}
\end{equation}
or, equivalently,
\begin{equation}
    \ddot {\bb u} =
    \bb H\, \nablaa \Fa
    + o(\varepsilon^4)
\end{equation}
We manipulate the Lagrangian \eqref{eq:Lagr} working on the action
\begin{equation}    
\begin{aligned}
    a=\int_{t_0}^{t_1}\int_{\Omega_0}\mathcal L \,\mathrm d\Omega_0\mathrm dt
    &=    \int_{t_0}^{t_1}\int_{\Omega_0}\bigg[
    \frac12\rho_0 |\bb v|^2
    - \rho_0\Big(h-\frac{1}{24}\big(\bb H\,\nablaa\Fa\big)\cdot\nablaa\Fa
    \Big)
     \bigg]\,\mathrm d\Omega_0\mathrm dt
    \\
    &\approx    \int_{t_0}^{t_1}\int_{\Omega_0}\bigg[
    \frac12\rho_0 |\bb v|^2
    +\frac{1}{24}\rho_0\,\ddot{\bb u}\cdot\nablaa\Fa   - \rho_0 h
    \bigg]\,\mathrm d\Omega_0\mathrm dt
    \\
    &=    \int_{t_0}^{t_1}\int_{\Omega_0}\bigg[
    \frac12\rho_0 |\bb v|^2
    +\frac{1}{24}\rho_0\,\nablaa\dot{\bb u}\cdot\dot{\bb F}\bb a   - \rho_0 h
    \bigg]\,\mathrm d\Omega_0\mathrm dt
    \\
    &= \int_{t_0}^{t_1}\int_{\Omega_0} \bigg[
    \frac12\rho_0 |\bb v|^2
    +\frac{1}{24}\rho_0\,\dot{\bb F}\bb a\cdot\dot{\bb F}\bb a   - \rho_0 h
    \bigg]\,\mathrm d\Omega_0\mathrm dt,
\end{aligned}
\end{equation}
where integration by parts in space and in time have been performed.
We explicit $h$ and obtain
\begin{align}\label{eq:lagr well posed}
    \mathcal L = \frac12\rho_0 |\bb v|^2
    +\frac{1}{24}\rho_0\,|\dot{\bb F}\bb a|^2   - \rho_0\frac{\kH}{n+1} \Big(|\bb a|-|\Fa|\Big)^{n+1}_+.
\end{align}
From this Lagrangian, we recognise the following two non-negative energy functions
\begin{align}\label{eq:effective K and V}
    \mathcal K &= \frac12\rho_0 |\bb v|^2
    +\frac{1}{24}\rho_0\,|\dot{\bb F}\bb a|^2,
    &
    \mathcal V &= \rho_0\frac{\kH}{n+1} \Big(|\bb a|-|\Fa|\Big)^{n+1}_+.
\end{align}

\subsection{Lagrangian of the 2D stack of grains}
All the results we have obtained so far recover the well-known one-dimensional chain of grains under the assumption of a one-dimensional motion, as one can quickly verify. Such a motion is described by a wave equation rich of features, the reader is referred to the existing theoretical, numerical and experimental literature on this topic \cite{nesterenko1994solitary,gavrilyuk1996media,coste1999validity,nesterenko2013dynamics}.

In two dimension, the interactions between particles is much more complex. We leverage the hypothesis that there is no coupling between the contacts from different directions and obtain the energies of the 2D stack by performing a sum:
\begin{equation}\label{eq:k and V sum}
    \mathcal K_\mathrm{2D} = \frac12\rho_0 |\bb v|^2 + \frac{1}{24}\rho_0
     \sum_{i=1}^3|\dot{\bb F}\bb a_i|^2,
\qquad
    \mathcal V_\mathrm{2D} =
     \rho_0\sum_{i=1}^3\frac{\kH}{n+1} \Big(|\bb a_i|-|\Fa_i|\Big)_+^{n+1},   
\end{equation}
where the Lagrangian vectors $\bb a_i$ connect the three pairs of neighbouring grains.
This formulation can be extended to random particle assemblies, as proposed, for example, in \cite{cambou1998micromechanical,liao2000pressure,muhlhaus1996dispersion,placidi2021micromechanics}. In these approaches, the effective potential energy is obtained by summing the contributions of all contact orientations, often represented through the so-called fabric tensor. Kinetic energy is modelled differently.
A fundamental assumption underlying such a homogenization procedure is that the contact state is statistically identical for all particles. This requirement is satisfied only when the distribution of contacts is locally centrosymmetric, so that each contact direction is balanced by its opposite. Note that this same assumption underlies the derivation presented in Section~\ref{sec:homogenization}, where the contact network is explicitly assumed to be symmetric.
Because of the dependency on $\dot{\bb F}$, the Lagrangian built with \eqref{eq:k and V sum} leads to a dispersive model. From this section on, we consider the effective model at the leading order and neglect such a term.

\section{Hyperbolicity of the leading-order model}
\label{sec:hyperbolicity}

At the leading order, the Lagrangian reads as
\begin{equation}
    \mathcal L = \tfrac12\rho_0|\bb v|^2-\rho_0\, e(\bb F),
\end{equation}
where the energy function $e(\bb F)$ specific to mass is
\begin{equation}\label{eq:e more contacts}
    e(\bb F) =
     \sum_{i=1}^3\frac{\kH}{n+1} \Big(|\bb a_i|-|\Fa_i|\Big)_+^{n+1}.
\end{equation}
Via the least action principle, we obtain the following equations of motion in the Eulerian frame:
\begin{equation}\label{eq:motion}
    \begin{dcases}
        \rho_t + \nablax\cdot(\rho\bb v) =0
        \\
        (\rho\bb v)_t + \nablax\cdot(\rho\bb v\otimes\bb v-\bs\sigma)=\bs 0
        \\
         (\rho\bb F)_t + (\bb v\cdot\nabla)(\rho\bb F) +\rho \big((\nablax\cdot\bb v)\; \bb I - \nablax\bb v\big)\bb F = \bs 0 
    \end{dcases}\;,
\end{equation}
where $\nablax$ is the differential operator in Eulerian coordinates and the Cauchy stress tensor is
\begin{equation}
    \bs\sigma \coloneqq \rho\bb F\frac{\partial e}{\partial\bb F}
    = -\kH\, \rho\sum_{i=1}^3 \frac{\big(|\bb a_i|-|\Fa_i|\big)^n_+}{|\Fa_i|}\;
    \Fa_i\otimes\Fa_i.
\end{equation}
Note that $\bs\sigma$ is symmetric
and it is composed by three uniaxial compressions aligned with the deformed lattice vectors $\Fa_i$.


To assess the stability of this model, we write the system~\eqref{eq:motion} in non-conservative form
\begin{equation}
    \bb U_t + \bb A(\bb U)\bb U_{x} +  \bb B(\bb U)\bb U_y =  \bb S(\bb U)
\end{equation}
and compute the characteristic speeds for the propagation along an arbitrary direction $\theta$ from the eigenvalues of $\bb A\cos\theta+\bb B\sin\theta$. 
Equivalently, we study the propagation along the $x$-axis from the spectral structure of $\bb A$, leaving all the vectors unspecified; we select different orientations of the coordinates to study particular  motions. For example, given a state of deformation, we will find counterexamples by assuming the $x$-axis to be aligned in a particular direction. The explicit formula of $\bb A$ is provided in Appendix~\ref{sec:app A}.

The three directions of contact are not always active, and the energy~\eqref{eq:e more contacts} simplifies accordingly.
In the following, we consider separately the cases in which one, two, or three contact directions are active, and investigate the hyperbolicity of the system of equations \eqref{eq:motion}.
The case in which no contact is active is not considered. Without contacts, the interaction energy vanishes identically and the continuum model loses its constitutive response, leaving the equations of motion underdetermined and the system weakly hyperbolic.
Such a situation corresponds to a dilute granular medium rather than the dense assemblies considered in the present work. The dynamics in this regime is more appropriately described by kinetic theories or continuum models developed for granular gases, and therefore lie beyond the scope of this study.

\subsection{One direction of contact}
\label{sec:1 contact}
\begin{figure}
    \centering
    \includegraphics[scale=0.8]{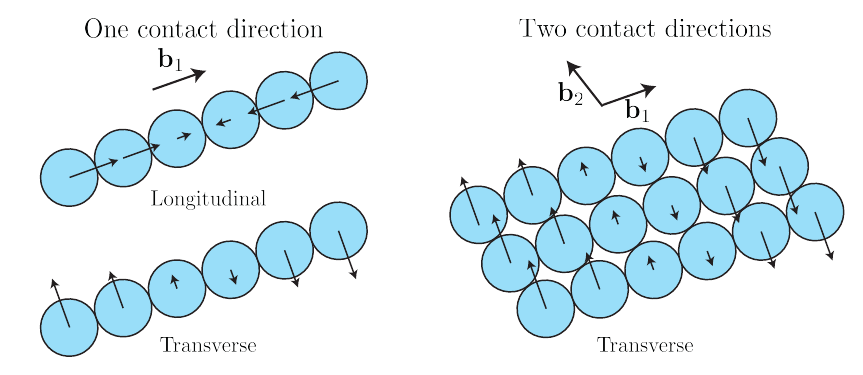}
    \caption{Left, longitudinal and transverse waves for one contact direction. Right, unstable transverse wave with two contact directions: the wave propagates orthogonally to $\bb b_2$ with a polarisation orthogonal to $\bb b_1$.
    }
    \label{fig:instability}
\end{figure}

We first assume that only one contact is active, that is the amount of compression along one direction is positive $\xi_1\coloneqq(|\bb a_1|-|\Fa_1|)_+>0$, and $\xi_i\coloneqq(|\bb a_i|-|\bb F\bb a_i|)_+=0$  for $i=2,3$.
This enables us to identify the similarities and differences with the classical one-dimensional model by Nesterenko.
In this case, the eigenvalues of $\bb A$ are
\begin{align}\label{eq:eigenvalues 1 cont}
    &\lambda_{1,2,3}=v_1,
    &
    &\lambda_{4,5}= v_1 \pm
    b^x_1\, \sqrt{
    n\,\kH\,\xi_1^{n-1}},
    &
    &\lambda_{6,7}= v_1\pm
    b^x_1\,  \sqrt{
    -\tfrac{\kH}{|\bb b_1|} \xi_1^n},
\end{align}
where $\bb b_i=\Fa_i$ are the deformed lattice vector.
The variables $a_i^x$, $a_i^y$, $b_i^x$, and $b_i^y$ are the components of $\bb a_i$ and $\bb b_i$, respectively, i.e.\ $\bb a_i=[a_i^x,a_i^y]^\top$ and $\bb b_i=[b_i^x,b_i^y]^\top$.
We now analyse the eigenvectors of $\bb A$.
Given the vector of states
\begin{equation}\label{eq:states vector}
    \bb U \coloneqq \{\rho, \rho v_1, \rho v_2, \rho F_{11}, \rho F_{12}, \rho F_{21}, \rho F_{22}\},
\end{equation}
the columns of the following matrix are the eigenvectors that correspond, in the same order, to the eigenvalues in Eq.~\eqref{eq:eigenvalues 1 cont}:
\begin{equation}\label{eq:eigenvectors}
\begin{bmatrix}
 1 & 0 & 0 & 1 & 1 & 1 & 1 \\
 v_1 & 0 & 0 & v_1+\sqrt{\kH\A_1} & v_1-\sqrt{\kH\A_1} & v_1+\sqrt{-\kH\B_1} & v_1-\sqrt{-\kH\B_1} \\
 v_2 & 0 & 0 & v_2+\frac{b^y_1}{b^x_1}\sqrt{\kH\A_1} & v_2-\frac{b^y_1}{b^x_1}\sqrt{\kH\A_1}& v_2-\frac{b^x_1}{b^y_1}\sqrt{-\kH\B_1} & v_2+\frac{b^x_1}{b^y_1}\sqrt{-\kH\B_1} \\
 \frac{b^x_1\A_1}{a^x_1(\A_1-\B_1)} & 0 & -a^y_1 & 0 & 0 & 0 & 0 \\
 0 & 0 & a^x_1 & 0 & 0 & 0 & 0 \\
 0 & -a^y_1 & 0 & F_{21}-\frac{b^y_1}{b^x_1}F_{11} & F_{21}-\frac{b^y_1}{b^x_1}F_{11} & \frac{b^x_1}{b^y_1}F_{11}+F_{21} & \frac{b^x_1}{b^y_1}F_{11}+F_{21} \\
 \frac{b^y_1\A_1}{a^y_1(\A_1-\B_1)} & a^x_1 & 0 & F_{22}-\frac{b^y_1}{b^x_1}F_{12} & F_{22}-\frac{b^y_1}{b^x_1}F_{12} & \frac{b^x_1}{b^y_1}F_{12}+F_{22} & \frac{b^x_1}{b^y_1}F_{12}+F_{22}
\end{bmatrix}
\end{equation}
where we have defined the following non-negative quantities
\begin{align}
    \A_i &\coloneqq     n (b_i^x)^2\,  \big(|\bb a_i|-|\bb b_i|\big)_+^{n-1}, &
    \B_i &\coloneqq \frac{(b_i^x)^2}{|\bb b_i|} \big(|\bb a_i|-|\bb b_i|\big)_+^n.
\end{align}
We focus on the eigenvectors related to the pairs $\lambda_{4,5}$ and $\lambda_{6,7}$.
The eigenvalues $\lambda_{4,5}$ are real. By inspection of the corresponding eigenvectors, the terms related to $U_2=\rho v_1$ and $U_3=\rho v_2$ (i.e.\ $v_1\pm\sqrt{\kH\A_1}$ and $v_2\pm\frac{b^y_1}{b^x_1} \sqrt{\kH\A_1}$, respectively) form a vector parallel to $\bb b_1$, when $\bb v=\bs0$. Thus, the motion is longitudinal with respect to the actual alignment of grains and affects the state of compression. A schematic is provided in Figure~\ref{fig:instability}.
This corresponds to the one-dimensional motion described in the literature of one-dimensional granular chains.
\\
The eigenvalues $\lambda_{6,7}$ are complex and are associated to the last two columns. The terms related to $U_2=\rho v_1$ and $U_3=\rho v_2$ (i.e.\ $v_1\pm\sqrt{-\kH\B_1}$ and $v_2\mp\frac{b^y_1}{b^x_1} \sqrt{-\kH\B_1}$, respectively) form a vector orthogonal to $\bb b_1$, when $\bb v=\bs0$. Thus, the motion is transverse with respect to the actual alignment of grains.
As this motion does not change the distance between two grains (not affecting the energy), the continuous effective model suffers from an instability problem: an arbitrary number of phase transition (shear bands in this case) can appear. This phenomenon is typically related to a non-convex energy, as occurs in the well-known Ericksen's bar \cite{ericksen1992introduction,bourgeois2020dynamics} and in the phase transition of fluids \cite{didierlaurent2025solving,dhaouadi2025first}.
In these scenarios, the instability is observed in one-dimensional problems or in volumetric expansion, where either flexural or longitudinal waves are unstable.
On the contrary, the model we propose simultaneously exhibits stable-longitudinal and unstable-transverse waves.

\subsection{Two directions of contact}
\label{sec:2 contacts}

We assume that two contacts are active: $\xi_i=(|\bb a_i|-|\Fa_i|)_+>0$ for $i=1,2$, and $\xi_3=0$. By denoting $\bb b_i=\Fa_i$, we obtain the following eigenvalues of the matrix $\bb A(\bb U)$:
\begin{equation}\label{eq:two contacts_1}
\begin{aligned}
    \lambda_{1,2,3} = v_1,
\quad
\lambda_{4,5,6,7}&=v_1\pm          \sqrt{\frac{\kH}{2}} 
        \Bigg\{
        \A_1-\B_1
        +\A_2-\B_2 
        \\[-6pt]
        \pm\bigg[
        (& \A_1+\B_1  - \A_2- \B_2 )^2
        + 4 (\A_1+\B_1) (\A_2+\B_2) (\hat{\bb b}_1\cdot\hat{\bb b}_2)^2
        \bigg]^{1/2}
        \Bigg\}^{1/2},
\end{aligned}
\end{equation}
where $\hat{\bb b}_i\coloneqq\bb b_i/|\bb b_i|$.
The nested square root is real, but we can always find a direction that makes the eigenvalues $\lambda_{4,5,6,7}$ complex.
Indeed, we can rotate the Eulerian coordinates such that the $x$-axis is orthogonal to $\bb b_2$ (i.e.\ $b_2^x=0$). Thus, $\A_2=\B_2=0$ and $\lambda_{4,5,6,7}$ are identical to the case with one direction of contact: a pair of eigenvalues is real and a pair is complex. Analogously to Section~\ref{sec:1 contact}, the polarisations of such motions are such that the particles move in the directions parallel or orthogonal to $\bb b_1$.
To help visualisation, a schematic of an unstable wave is provided in Figure~\ref{fig:instability}. The dual reasoning applies by considering a wave that propagates orthogonally to $\bb b_1$ (i.e.\ by rotating the Eulerian axis such that $b_1^x=0$).

Nevertheless, the interplay between the two contact directions \emph{helps} hyperbolicity; this is more evident by looking at the particular configuration where $\bb b_1$ and $\bb b_2$ are orthogonal. 
This configuration corresponds to the linearized kinematics studied by Leonard \emph{et al.}~\cite{leonard2013directional}.
We obtain
\begin{equation}\label{eq:two contacts_3}
    \lambda_{1,2,3} = v_1,
    \quad
     \lambda_{4,5}=v_1\pm  \sqrt{\kH}\sqrt{\A_1 - \B_2},
     \quad
    \lambda_{6,7}=v_1\pm \sqrt{\kH}\sqrt{\A_2 - \B_1}.
\end{equation}
The contacts along the two directions couple: considering $\lambda_{4,5}$, the unstable motion due to the contacts along $\bb b_2$ (i.e.\ the motion orthogonal to $\bb b_2$) is balanced by the stable motion due to the contacts along $\bb b_1$ (i.e.\ the motion parallel to $\bb b_1$). Depending on whether or not $\A_1$ is larger than $\B_2$, the two effects may result in stable or unstable motion. The dual reasoning holds for $\lambda_{6,7}$.
%
In particular, if the contacts are similar in all directions, i.e.\ $\xi_1\approx\xi_2$, and are of moderate intensity, $|\bb b_i|\gtrsim|\bb a_i|$, we have that $\A_i\gg\B_j,\, \forall i,j$ and the eigenvalues $\lambda_{4,5,6,7}$ are real.
Conversely, if the contacts have sufficiently different or large intensities, we may have two or four unstable waves.

Finally, the eigenvectors in case of two contact directions are obtained as a particular case of the three contact directions shown in Appendix~\ref{sec:app A} and discussed below.


\subsection{Three directions of contact}
\label{sec:3 contacts}

We assume now that three contacts are active: $|\bb a_i|-|\Fa_i|>0$, for $i=1,2,3$. The eigenvalues are:
\begin{equation}\label{eq:eigenvalues 3 cont}
\begin{aligned}
\lambda_{1,2,3} = v_1,
\qquad
    &\lambda_{4,5,6,7}=v_1\pm          \sqrt{\frac{\kH}{2}} 
        \Bigg\{
        \sum_i\big(\A_i -\B_i)
        \\[-6pt]
        &\pm\bigg[
        \Big(
         \sum_i\big(\A_i +\B_i)
        \Big)^2
        -4\sum_{i<j} (\A_i+\B_i)(\A_j+\B_j) |\hat{\bb b}_i\times\hat{\bb b}_j|^2
        \bigg]^{1/2}
        \Bigg\}^{1/2}.
\end{aligned}
\end{equation}
This formula is valid for one, two and three contact directions, suggesting that it may naturally extend to a broader class of energies of the form given in Eq.~\eqref{eq:e more contacts}.

\noindent
The argument of the nested square root in Eq.~\eqref{eq:eigenvalues 3 cont} is non-negative. To prove this, we write it as
\begin{equation}
    \Big[\sum_i\Big(1
    -2 \frac{(b_i^y)^2}{|\bb b_i|^2} \Big)(\A_i + \B_i)
    \Big]^2
    +4 \Big(\sum_i\frac{b_i^x b_i^y}{|\bb b_i|^2} (\A_i + \B_i) \Big)^2 \ge 0.
\end{equation}
To establish hyperbolicity, it is necessary to determine the conditions under which the outer square root is real-valued. Owing to the complexity of these expressions, we restrict our analysis to a range of isotropic deformations. For all numerical simulations presented in Section~\ref{sec:numerics}, we have verified \emph{a posteriori} that the eigenvalues remain real at every time step.
Let us assume the three directions of contact are active ($\xi_i>0$, for $i=1,2,3$) and that they are oriented at an angle of $\pi/3$ to each other:
\begin{equation}
    |\hat{\bb b}_1\times\hat{\bb b}_2|^2 =
    |\hat{\bb b}_2\times\hat{\bb b}_3|^2 =
    |\hat{\bb b}_3\times\hat{\bb b}_1|^2 = \frac34.
\end{equation}
Without loss of generality, we set $\alpha_1=\theta$, $\alpha_2=\theta+\frac\pi3$, and $\alpha_3=\theta+\frac23\pi$. In this way, we explore different states of deformation by varying $\theta$. This is equivalent to choosing $\alpha_1=0$, $\alpha_2=\frac\pi3$, and $\alpha_3=\frac23\pi$ and changing the direction $\bb e^x$ of the $x$-axis as $\bb e^x=[\cos\theta,\sin(-\theta)]^\top$.
We also assume that $|\bb a_1|=|\bb a_2|=|\bb a_3|=|\bb a|$ and $|\bb b_1|=|\bb b_2|=|\bb b_3|=|\bb b|$ (thus, $\xi_1=\xi_2=\xi_3=\bb \xi$). Note that $\A_i=n \cos^2(\alpha_i)\,|\bb b|^2\,\xi^{n-1}$ and $\B_i=\cos^2(\alpha_i)\, |\bb b| \xi^n$. This corresponds to an isotropic deformation, i.e.\ to $\bb F\propto\bb I$.
Under these hypotheses, the eigenvalues $\lambda_{4,5,6,7}$ are written as:
\begin{equation}\label{eq:isotropic speeds}
    \lambda_{4,5,6,7}=v_1\pm          \frac12|\bb b| \sqrt{3\kH 
         \Big(n+\frac{\xi}{|\bb b|}\Big)\xi^{n-1}  }
        \sqrt{ 1\pm\frac12-\frac{2\xi}{n|\bb b|+\xi} }\;.
\end{equation}
Remarkably, they are independent of the orientation of the grains (i.e.\ no dependence on $\theta$), and the effective material is isotropic.
When the deformation of the particles is moderate, we have that $|\bb a|\sim|\bb b|$, and so $\xi\ll |\bb b|$ and the eigenvalues are real. For large deformations, two or four eigenvalues become complex; this and the case discussed above with only two orthogonal active contact directions remark the fail of Cauchy--Born hypothesis under large deformations. Such instabilities resemble those discussed by Friesecke (2002) \cite{friesecke2002validity} (see Fig.~7 for a clear schematic).\\
Notably, we recover the results that Chong \emph{et al. }\cite{chong2016conical} obtained by analysing the dispersion of a prestressed hexagonal packing: for $\xi / |\bb a| \le 1/3$, the eigenvalues are real; for $1/3<\xi / |\bb a| \le 9/11$, two eigenvalues are complex and two real; and for $\xi / |\bb a| >9/11$ four eigenvalues are complex.
The unstable situations are unlikely in practice. For example, by assuming $\A_1=\A_2$ and $\B_1=\B_2$, the eigenvalues are complex when $\A_i<\B_i$ that corresponds to $|\bb b_i|<|\bb a_i|/(n+1)=0.4|\bb a_i|$. This is an extreme state of compression, that would also render Hertz's law inapplicable.

Finally, the eigenvectors of the Jacobian $\bb A(\bb U)$ are computed symbolically using Wolfram Mathematica, with their analytical expressions provided in the appendix. Although proving their linear independence is challenging, they were verified to be linearly independent numerically in all simulations conducted in this study.

\subsection{Summary about the hyperbolicity}

In summary, the hyperbolicity analysis leads to the following conclusions:
\begin{description}[topsep=2pt, itemsep=2pt, leftmargin=20pt]
    \item[One contact direction:] The system always exhibits two hyperbolic and two elliptic wave modes, independently of the deformation state and the propagation direction.
    
    \item[Two contact directions:] There always exist two propagation directions along which the model reduces to the one-dimensional behaviour, with two hyperbolic and two elliptic modes. For other propagation directions, the stability depends on the contact forces. In particular, sufficiently strong contacts may lead to either two or four elliptic modes.
    
    \item[Three contact directions:] No propagation direction is intrinsically equivalent to the one-dimensional case. Owing to the complexity of the governing equations, the formal analysis has been restricted to isotropic deformations. In this setting, the system remains hyperbolic except under extremely strong contact forces. More generally, the space of admissible deformations is expected to contain both hyperbolic and non-hyperbolic regions, although a complete characterization lies beyond the scope of the present work. The Jacobian eigenvectors are computed symbolically.
\end{description}
In the numerical examples presented in the following section, hyperbolicity is assessed \emph{a posteriori} by verifying that the eigenvalues of the governing equations remain real throughout the simulations. This approach is justified because the deformation stays close to the isotropic compression state analysed above, where the hyperbolicity has been characterised.
\\
If the leading-order model loses hyperbolicity, the higher-order energy terms in \eqref{eq:effective K and V} may regularise the dynamics through dispersion.

\section{Numerical results}
\label{sec:numerics}

\begin{figure}
    \centering
    \includegraphics[width=0.825\linewidth]{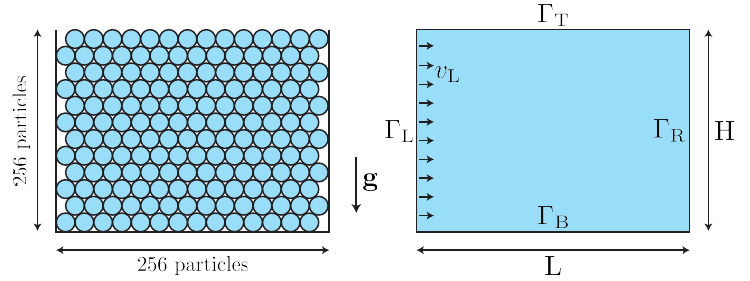}
    \caption{Left: setup of the discrete simulations considered by Melin~\cite{melin1994wave}; right: setup of the corresponding simulations with our continuous model.
    }
    \label{fig:nume scheme}
\end{figure}

To assess the predictive capabilities of the proposed continuum model, we consider the benchmark configuration introduced by Melin~\cite{melin1994wave}, originally investigated using the Discrete Element Method (DEM). A schematic of the computational setup is shown in Figure~\ref{fig:nume scheme}. According to the experimental setting, we consider a granular assembly composed of $256 \times 256$ identical circular particles arranged in a hexagonal packing and placed in the vertical plane. The left, right, and bottom boundaries are fixed, while the top boundary is assumed to be stress free. There is no friction between the grains or between the grains and the walls.
\\
The assembly is first allowed to settle under the action of gravity. This settling process generates a depth-dependent stress field and, consequently, a spatial gradient in the wave speed. Remarkably, existing effective models do not account for the simultaneous presence of contact and geometric nonlinearities \cite{gilles2003low,hua2019wave,merkel2010elastic,chong2016conical,awasthi2012propagation,leonard2012stress,leonard2013directional}.
Once static equilibrium is reached, wave propagation is investigated by imposing a time-varying uniform displacement at the left boundary. The resulting disturbance propagates through the granular assembly and is analysed by comparing the predictions of the continuum model with the corresponding DEM simulations.
\\
As no analytical solutions are available for the problems under analysis, we resort to numerical methods. The continuous model is simulated using a finite-volume scheme with explicit time integration. This approach is particularly efficient for large granular assemblies compared with discrete simulations, while retaining the ability to capture the relevant macroscopic wave-propagation phenomena within the continuum framework.

\subsection{Numerical solution of the effective model via Finite Volumes}

The system to be solved is
\begin{subequations}\label{eq:motion num}
    \begin{numcases}{}
        \rho_t + \nablax\cdot(\rho\bb v) = 0
        \label{eq:motion num-a}
        \\
        (\rho\bb v)_t + \nablax\cdot(\rho\bb v\otimes\bb v-\sigma)=\rho\bb g - \gamma\rho\bb v
        \hspace{70pt}.
        \label{eq:motion num-b}
        \\
        (\rho\bb F)_t + \nablax\cdot(\rho\bb F\otimes\bb v-\rho\bb v\otimes\bb F^\top) = 
        -\bb v\otimes\nablax\cdot(\rho\bb F^\top)
        \label{eq:motion num-c}
    \end{numcases}
\end{subequations}
This quasi-conservative form is required by the numerical schemes discussed below and is obtained by manipulation of Eq.~\eqref{eq:motion}
as done, for example, by Barton \emph{et al.}~\cite{barton2010eulerian}. Remarkably, the source term of Eq.~\eqref{eq:motion num-c} is zero for smooth solutions due to the Piola's constraint $\nablax\cdot(\rho\bb F^\top) = \bs 0$.
We recognise the hyperbolic part of the system~\eqref{eq:motion num} as that composed by the terms on the left-hand side. Notably, its eigenvalues and eigenvectors are different from that discussed in Section~\ref{sec:hyperbolicity}, and a closed-form spectral decomposition is more complicated, as discussed in Appendix~\ref{sec:app B}.
\\
Compared with Eq.~\eqref{eq:motion}, two additional terms are included on the right-hand side of the momentum balance: the gravitational force $\rho\bb g$, with gravity $\bb g=[0,\SI{-9.81}{\meter\per\second\squared}]^\top$, and the viscous term $-\gamma\rho\bb v$, with damping coefficient $\gamma$. The value of the latter is adjusted according to the simulation being performed. In a physical experiment, where the particles are confined to move in a plane between two parallel plates, this damping term can be interpreted as a Maxwell-type friction arising from the interaction between the particles and the confining plates. In the following, friction is only used to compute the steady-state solution due to the effect of gravity.

We consider a rectangular domain with impenetrable, slip conditions on its boundaries:
\begin{equation}
    \begin{dcases}
        \bb v\cdot\bb n = v_\mathrm{L}(t) &\text{and}\quad (\bb I-\bb n\otimes\bb n)\bs\sigma\bb n=\bs0 \quad \text{on } \Gamma_\mathrm{L}
        \\
        \bb v\cdot\bb n = 0 &\text{and}\quad (\bb I-\bb n\otimes\bb n)\bs\sigma\bb n=\bs0 \quad \text{on } \Gamma_\mathrm{R} \cup \Gamma_\mathrm{B} \cup \Gamma_\mathrm{T}
    \end{dcases},
\end{equation}
where $\bb n$ is the outgoing normal at the boundary, and the velocity $v_\mathrm{L}(t)$ is imposed on the left boundary; the function $v_\mathrm{L}$ is chosen according to the simulation we consider.
Finally, we choose the following initial conditions:
\begin{align}
\rho|_{t=0}&= \rho_{0,\mathrm{hex}} (1+\epsilon), &
\bb v|_{t=0}&=\bs0, &
\bb F|_{t=0}&=\mathrm{diag}\big[(1+\epsilon)^{-0.1}, (1+\epsilon)^{-0.9}\big],
\end{align}
and $\bb a_1 =[1,0]^\top R$, $\bb a_2 =[\sqrt3/2,1/2]^\top R$, $\bb a_3 =[\sqrt3/2,-1/2]^\top R$, where $R$ is the particle radius.
This corresponds to the discrete configuration shown in Figure~\ref{fig:schematic}, in which the particles are arranged in a hexagonal lattice. To prevent loss of contacts, the simulation is initialized with a small anisotropic precompression given by
\cbox{$\epsilon=\SI{2.15e-4}{}$} distributed along $x$ and $y$ according to the exponents $0.1$ and $0.9$. This choice corresponds to a compression along $y$ that will spread due to gravity and a smaller compression along $x$, that adds a margin to avoid the loss of contact of the particles.
The value of $\epsilon$ has been tuned by trial and error so that the stress $\sigma_{yy}$ at $\Gamma_\mathrm{T}$ becomes close to $0$ after the settling due to gravity, approximating a no-traction condition.

The nonlinear system is solved using an in-house finite-volume solver implemented in \Matlab{}, which is available via the open repository referred to at the end of the paper. The governing equations are discretized on a cell-centred grid in terms of the conservative variables introduced in Eq.~\eqref{eq:states vector}. 
The gravity source terms of the momentum equation \eqref{eq:motion num-b} are expected to be mainly balanced by $\nabla\cdot\bs\sigma$. This is especially the case when the state is about its static equilibrium, for which $\bb v\approx\bs0$. Numerical solutions of such situations are known to suffer from the so-called well-balancing problem. The existing literature aiming to fix this problem mainly focuses on equations for gravity waves in fluids. In such a setting, the static equilibrium can be computed in closed form and the numerical scheme can be specifically designed to preserve it. However, no analytical equilibrium solution is available for the model proposed in this paper. For this reason, we adapt the flux-splitting approach proposed by Bale \emph{et al.}~\cite{bale2003wave} with a Godunov splitting . The technical details are provided in the appendix.


Due to the lack of details in Melin \cite{melin1994wave}, similar parameters are considered when possible and the numerical experiments are compared only from a qualitative point of view.
We choose particles of radius $R=\SI{5}{\milli\meter}$ with physical properties:
\begin{align}
    \rho_\mathrm{p}&=\SI{7650}{\kilo\gram\per\cubic\meter},&
    E_\mathrm{p}&=\SI{226}{\giga\pascal},&
    \nu_\mathrm{p}&=\SI{0.3}{}.
\end{align}
The computational domain is set to $[0,L]\times[0,H]$, where $L=512 R=\SI{2.56}{\meter}$ and $H=256\sqrt3R=\SI{2.22}{\meter}$, and it is discretised using \cbox{$200\times173$} cells with size $(\mathrm dx,\mathrm dy)=(\SI{12.8}{\milli\meter},\SI{12.8}{\milli\meter})$. Two layers of ghost cells are added on each boundary.
The numerical values are initialised at t=0 using Eq.~(4.3) and the settling simulation is run.

\subsection{Settling under gravity}

To accelerate the settling process, the friction coefficient is set to $\gamma=\SI{800}{\per\second}$. This value was determined through empirical iterations aimed at a quick settling. As in a linear spring--mass--damper system, this value brings the system close to critical damping, suppressing oscillations and promoting an exponential decay of the energy.
The boundary velocity is prescribed as $v_\mathrm{L}(t)=\SI{0}{\meter\per\second}$. Under the action of gravity, the material falls and undergoes compression.
The simulation is run for a simulated time of \SI{50}{\milli\second}, which is sufficient to dissipate most of the energy. At this point, the damping is removed by setting $\gamma=\SI{0}{\newton\second\per\meter}$, and the simulation is continued till $t=\SI{100}{\milli\second}$ to assess whether the numerical scheme preserves the well-balanced state or develops spurious oscillations. The simulation was performed on a laptop equipped with an Intel Core i9-13950HX processor and \SI{32}{\giga\byte} of RAM. The computation required 4046 time steps, each requiring an average computational time of approximately $\SI{0.44}{\second}$, for a total of approximately \SI{1798}{\second}.

\begin{figure}
    \centering
    \subfloat[]{\includegraphics[scale=0.55]{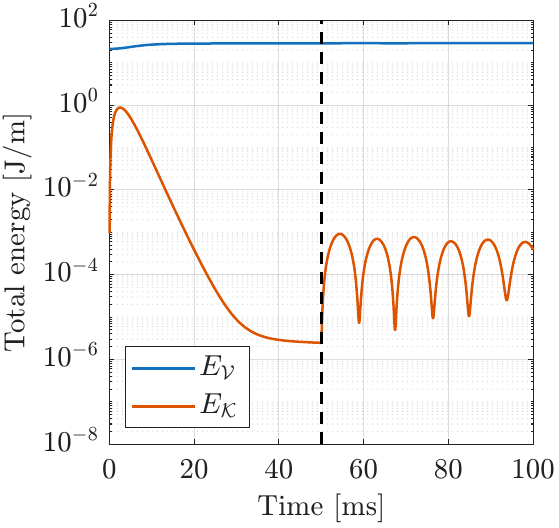}
    \label{fig:energy-a}}\;
    \subfloat[]{\includegraphics[scale=0.55]{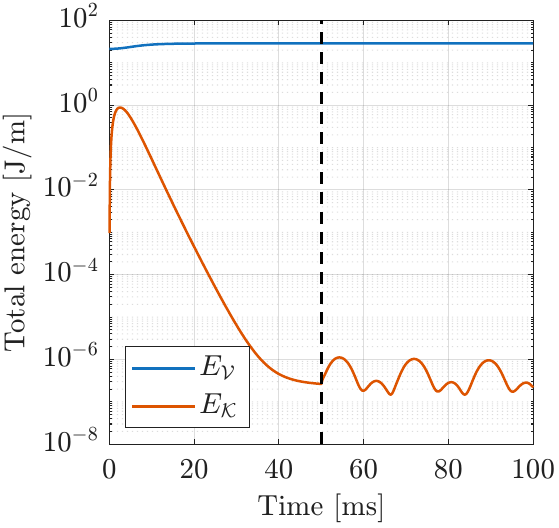}
    \label{fig:energy-b}}
    \caption{Total energies during the settling (a) without and (b) with the well-balanced scheme.
    }
    \label{fig:energy}
\end{figure}

\begin{figure}
    \centering
    \subfloat{\includegraphics[scale=0.52,trim=0 30 0 0,clip]{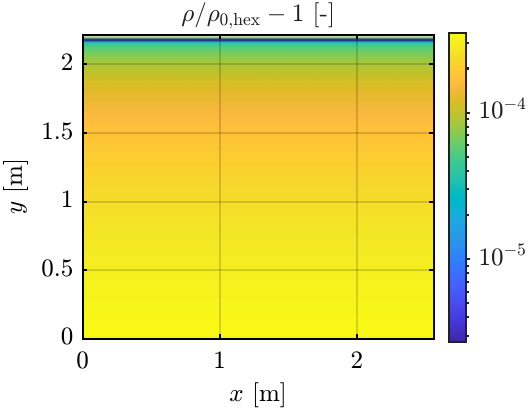}}\;
    \subfloat{\includegraphics[scale=0.52,trim=35 30 0 0,clip]{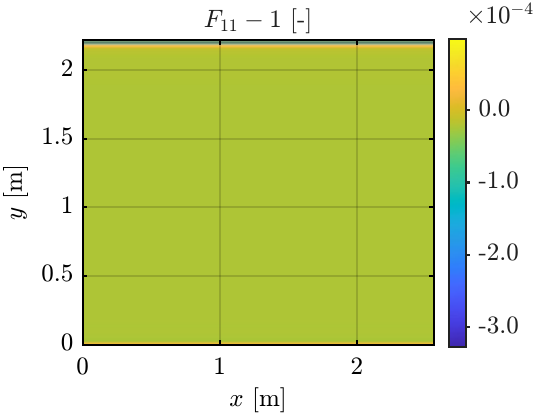}}\;
    \subfloat{\includegraphics[scale=0.52,trim=35 30 0 0,clip]{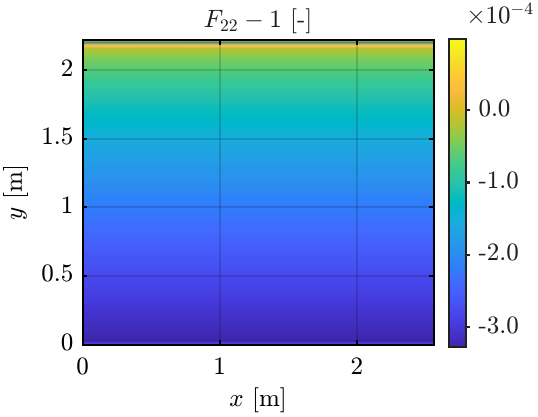}}
    \\[-10pt]
    \subfloat{\includegraphics[scale=0.52,trim=0 0 0 0,clip]{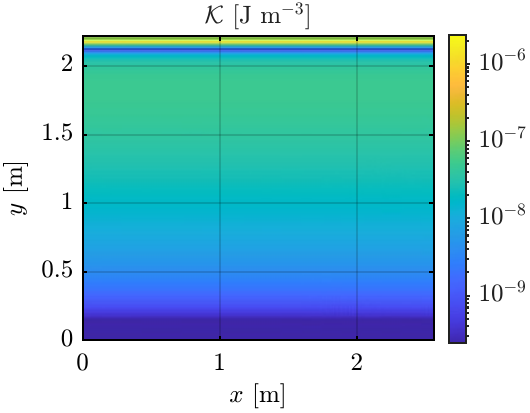}}\;
    \subfloat{\includegraphics[scale=0.52,trim=35 0 0 0,clip]{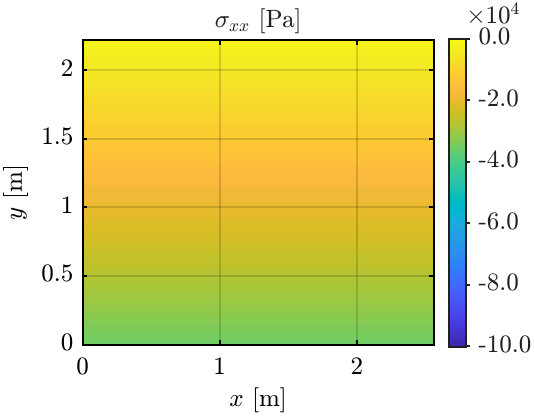}}\;
    \subfloat{\includegraphics[scale=0.52,trim=35 0 0 0,clip]{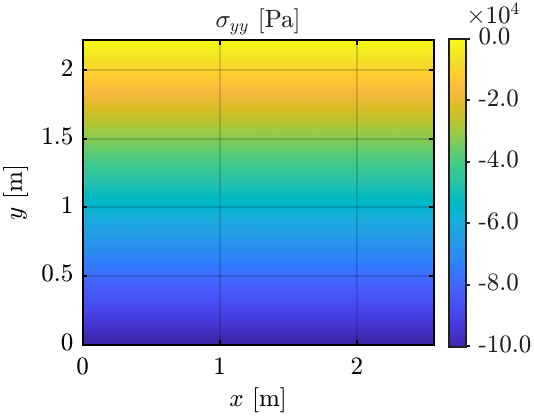}}
    \caption{Field variables after the settling under gravity at $t=\SI{50}{\milli\second}$. The values of $F_{12}$, $F_{21}$ and $\sigma_{xy}$ are zero up to the level of numerical precision and have not been displayed.
    }
    \label{fig:settling}
\end{figure}

Figure~\ref{fig:energy} shows the evolution of the total energies of the system over time.
For illustrative purposes, the energy shown in Figure~\ref{fig:energy-a} corresponds to a solution obtained without a well-balanced scheme: 
the system approaches a steady state equilibrium, but it oscillates as soon as the friction is removed.
Conversely, Figure~\ref{fig:energy-b} shows the energy obtained with the dedicated scheme. In this case, the kinetic energy is lower and the oscillations are much less important. More details are provided in the appendix. From now on, we consider only solutions obtained with the well-balanced scheme. 
The kinetic energy $E_\mathcal{K}\coloneqq\int_\Omega \mathcal K \mathrm d\bb x$ decreases steadily, reaching approximately \cbox{\SI{2.7e-7}{\joule\per\meter}} at $t=\SI{50}{\milli\second}$ and the energy stored in the contacts $E_\mathcal{V}\coloneqq\int_\Omega \mathcal V \mathrm d\bb x$ slightly increases. After $t=\SI{50}{\milli\second}$, limited oscillations in the kinetic energy are observed, indicating that the scheme successfully preserves the static equilibrium.
\\
The state obtained at the end of the settling process is shown in Figure~\ref{fig:settling}. In particular, the kinetic energy remains very small throughout the domain, but does not reach exactly zero due to numerical error. It has been observed that the maximum velocity decreases at a second-order rate with respect to the mesh size.
The stress $\sigma_{yy}$ at the top of the domain is approximately \cbox{\SI{460}{\pascal}}, which is much smaller than the value at the bottom, approximately \SI{1e5}{\pascal}.
After the compression, the stress $\sigma_{xx}$ is significantly smaller than $\sigma_{yy}$, but increases with the depth. This is consistent with the observations by Melin. In fact, the contact of the grains along the horizontal direction is weak.

\textbf{Wave velocities.}
Under gravity, the material is expected to develop a pressure gradient with depth, leading to a depth-dependent wave speed. This behaviour is well documented in both theoretical and experimental studies, see e.g.\ \cite{goddard1990nonlinear,jia2020sound,makse2004granular}.
We can compare the characteristic velocities of our model.
Assuming an isotropic compression within the validity range of Hertzian contact, one has $|\bb b|\lesssim|\bb a|$ and ${\xi=|\bb a|-|\bb b|\ll|\bb b|}$. The characteristic velocities given in Eq.~\eqref{eq:isotropic speeds} reduce to
\begin{equation}
    \lambda_{4,5,6,7}\approx v_1\pm          \frac12|\bb a| \sqrt{3\kH 
         n\,\xi^{n-1}  }
        \sqrt{ 1\pm\frac12 }=v_1\pm c_{\mathrm L,\mathrm T}.
\end{equation}
where $c_\mathrm{L}=\sqrt3\, c_\mathrm{T}$ are two sound speeds that, in analogy with solids, can be associated to longitudinal and transverse waves. Due to its invariance under rotation, the same expression is found considering any direction of propagation.
Considering Hertz's law, the confining contact force is $F\propto\xi^n$, thus
\begin{equation}\label{eq:veloc isotrop}
    c_{\mathrm L,\mathrm T}  \propto
    F^{\frac{n-1}{2n}} = F^{\frac16}
\end{equation}
recovering the well-known $F^{1/6}$ scaling.
This agreement is expected provided that the granular assembly consists of perfect spheres arranged in a regular contact network.
In fact, as discussed by Gilles and Coste \cite{gilles2003low}, at low confining pressures the contact forces become sufficiently small, so that surface roughness and geometrical imperfections significantly affect the contact mechanics. As a result, experiments often exhibit an alternative scaling close to $F^{1/4}$, that have never been reproduced by a model based on regular particles.
We note that, according to Eq.~\eqref{eq:veloc isotrop}, we obtain that both longitudinal and transverse waves follow the $F^{1/6}$ scaling.

\begin{figure}
    \centering
    \includegraphics[scale=0.4]{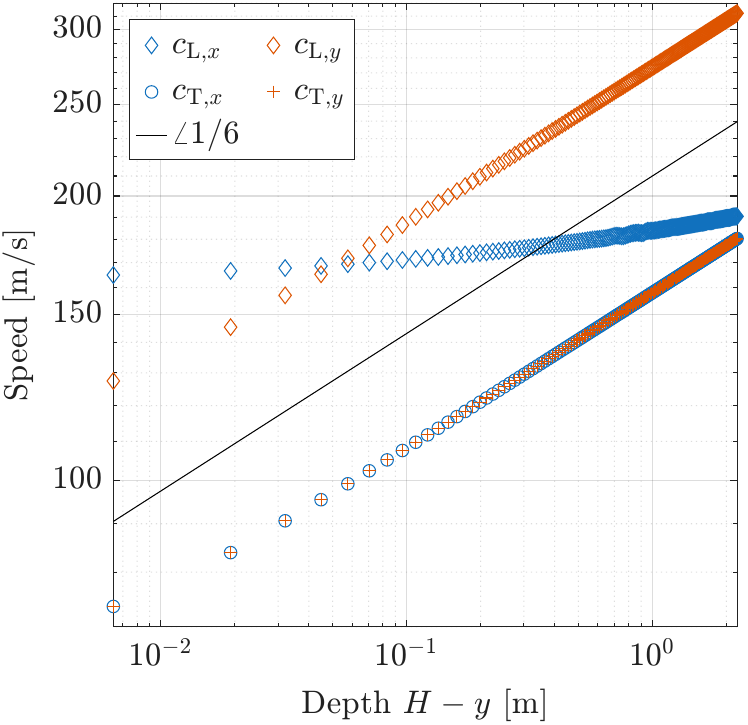}
    \caption{
    Comparison of the velocity gradients induced by gravity with the slope at $1/6$.
    }
    \label{fig:velocities}
\end{figure}

The state reached after the settling is far from an isotropic deformation, thus we use the eigenvalues in \eqref{eq:eigenvalues 3 cont}, that provide closed-form expressions for the wave speeds $c_{\mathrm{L,T}}$ for a general deformation state. We evaluate these expressions at the discrete simulation points along the vertical direction, since all state variables are constant along the horizontal axis.
Figure~\ref{fig:velocities} shows the four wave speeds, $c^x_{\mathrm{L,T}}$ and $c^y_{\mathrm{L,T}}$, as functions of the depth $H-y$, on a log--log scale. The differences between the horizontal and vertical wave speeds reflect the anisotropic deformation, which invalidates the isotropic approximation underlying Eq.~\eqref{eq:veloc isotrop}.
Based on the states in Figure~\ref{fig:settling}, isotropy is approximately recovered at $H-y=\SI{0.05}{\meter}$, where the wave speeds shown in Figure~\ref{fig:velocities} satisfy $c^x_{\mathrm{L,T}}=c^y_{\mathrm{L,T}}$.
The horizontal longitudinal wave speed, $c^x_{\mathrm{L}}$, is the velocity measured by Melin through the wave time-of-flight. As in his results, we observe a slope significantly smaller than $1/6$. A quantitative comparison is not possible, however, because the available information is insufficient to reproduce the experimental conditions and numerical values.
Conversely, the longitudinal speed along the vertical direction $c^y_\mathrm{L}$ has slope $1/6$.
\\
Finally, the two transverse wave speeds, $c^x_{\mathrm{T}}$ and $c^y_{\mathrm{T}}$, have slope $1/6$ and are nearly identical. The latter follows directly from \eqref{eq:eigenvalues 3 cont} by assuming that the angles between the vectors $\bb b_i$ are $\pi/3$, which is a good approximation of this settled configuration.
These speeds are difficult to measure in practice and our formulas propose the first analytical method for computing them.

\subsection{Wave propagation}

\begin{figure}
    \centering
    \subfloat{\includegraphics[scale=0.475,trim=0 30 45 0,clip]{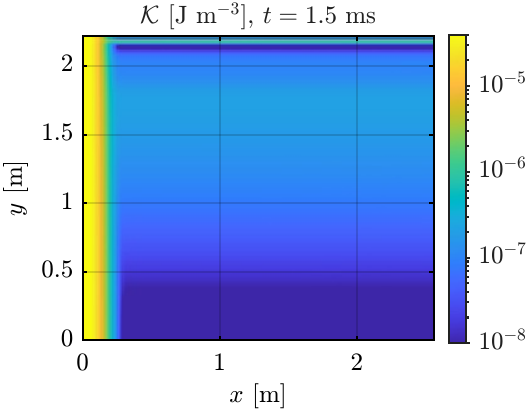}}\;
    \subfloat{\includegraphics[scale=0.475,trim=35 30 45 0,clip]{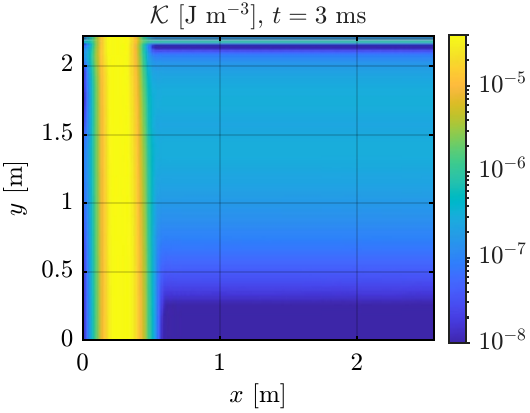}}\;
    \subfloat{\includegraphics[scale=0.475,trim=35 30 45 0,clip]{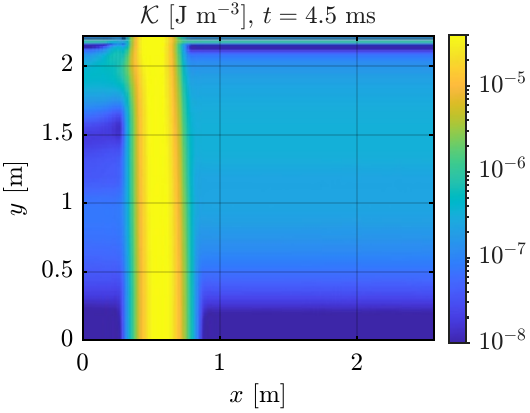}}\;
    \subfloat{\includegraphics[scale=0.475,trim=35 30 0 0,clip]{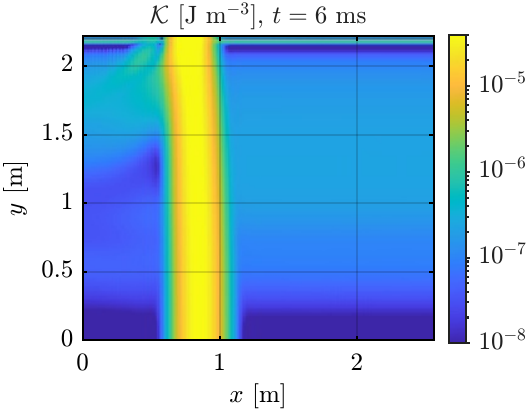}}
    \\[-10pt]
    \subfloat{\includegraphics[scale=0.475,trim=0 0 45 0,clip]{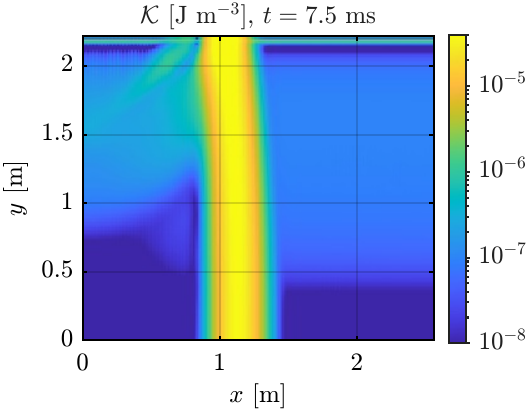}}\;
    \subfloat{\includegraphics[scale=0.475,trim=35 0 45 0,clip]{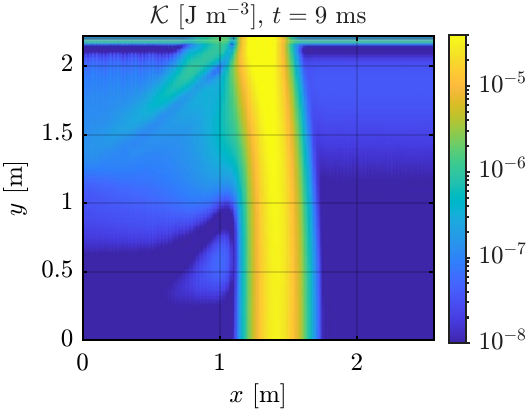}}\;
    \subfloat{\includegraphics[scale=0.475,trim=35 0 45 0,clip]{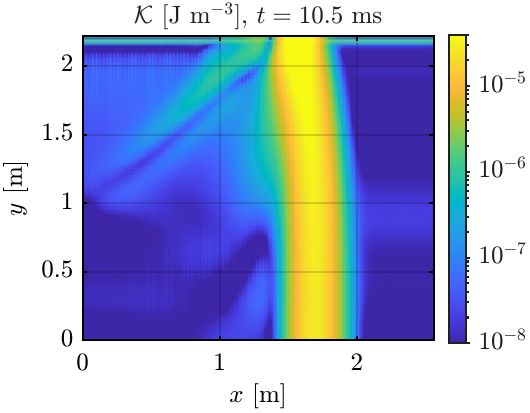}}\;
    \subfloat{\includegraphics[scale=0.475,trim=35 0 0 0,clip]{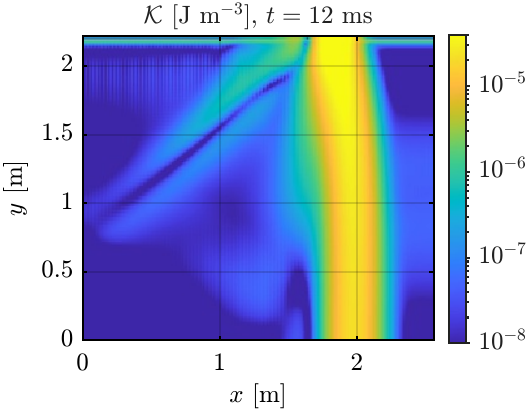}}
    \caption{Kinetic energy at different time instants.
    }
    \label{fig:kin}
\end{figure}

As a final numerical test, we simulate the propagation of a wave in the settled configuration, by setting the initial condition with the values obtained above at $t=\SI{50}{\milli\second}$.
The friction coefficient is set to $\gamma=\SI{0}{\newton\second\per\meter}$, and the boundary velocity $v_\mathrm{L}$ is prescribed as a Gaussian pulse,
\begin{equation}
    v_\mathrm{L}(t) = \frac{u_0}{w\sqrt{2\pi}}  \exp{-\frac12\Big(\frac{t-t_0}{w}\Big)^2};
\end{equation}
which corresponds to a step-like displacement of amplitude $u_0$. The pulse width and time delay are chosen as $w=\SI{0.5}{\milli\second}$ and $t_0=3w$, respectively. The value of $w$ is selected to be short enough for the imposed displacement to be effectively sudden, yet sufficiently larger than the characteristic simulation time step (approximately \cbox{\SI{0.025}{\milli\second}} for the parameters considered) so that the boundary velocity is accurately resolved in time.
The amplitude $u_0$ is set to \SI{10}{\percent} of the radius, i.e.\ $u_0=\SI{10}{\percent}R$. This is close to the numerical experiments performed by Melin.
\\
In Figure~\ref{fig:kin}, we report the kinetic energy at different time intants till $t=\SI{12}{\milli\second}$, shortly before the wave reaches the right wall and is reflected back. Due to the small amplitude of the displacement, the wave is mostly visible through the kinetic energy $\mathcal{K}$. The gradient of the wave velocity with depth causes the wave front to become slightly tilted.
The resulting wave profile is similar to that reported by Melin, with a coherent wave propagating towards the right boundary.
The top boundary, set to be rigid due to numerical difficulties, affects the motion nearby in a slightly different way from the discrete simulations by Melin. A more accurate treatment of this boundary requires further work.

\section{Conclusions}
\label{sec:conclusions}

In this study, we developed a nonlinear continuum model to describe the dynamics of two-dimensional regular granular assemblies. Starting from a discrete contact network of spherical particles, we derived an effective medium model up to second order using a long-wavelength homogenization procedure. By performing homogenization of the energy, the second-order expansion concerned only the Hertz-type contact nonlinearities between grains, leaving intact any geometric nonlinearity.
The resulting theory is a second-gradient hyperelastic continuum. Using the Boussinesq trick, we derive a bounded energy for the resulting constitutive law and formulate a physically meaningful model.
We investigated the hyperbolicity of the model at leading order and showed that at least three distinct contact directions are required for well-posedness. We obtained the eigenvalues and eigenvectors analytically, and discussed the different modes supported by the medium.
On the one hand, the model recovers results reported in the literature for hexagonal and cubic packings, as well as for different compression states and both stable and unstable configurations.
On the other hand, the analytical spectral decomposition enables us to derive explicit expressions for the propagation velocity. These expressions can distinguish between longitudinal and transverse modes, and account for the direction of propagation relative to the underlying contact network. This directional dependence highlights an important aspect of granular wave propagation, which is captured by the proposed homogenized model in the form of anisotropy.
\\
To solve the dynamics of the model, we have developed an efficient numerical strategy based on an explicit, second-order finite-volume discretisation. The scheme is well balanced, which is particularly important in the presence of gravity, and allows the equilibrium state to be preserved accurately while resolving finite-amplitude perturbations.
From the effective model, we recover the well-known power-law dependence of wave speed on depth, with an exponent of $1/6$ using the analytical expressions for isotropic compression. However, preliminary numerical simulations of gravitational settling reveal a non-uniform stress distribution, leading to a more complex dependence of wave speeds on depth that departs from the $1/6$ power law. This difference is consistent with previous Discrete Element Method simulations.

Several directions remain open for future work.
First, the present formulation provides a basis for extending the theory to more realistic granular packings, for instance by considering different minimal cells and contact-network topologies.
The restriction to three contact directions could be relaxed by constructing the energy from a larger set of contact directions. This would allow the model to capture more complex packing geometries while preserving its continuum and second-gradient structure.
Second, the model could be enriched to account for additional physical effects. In particular, friction could be described beyond the Maxwell-type friction considered here. The contact laws could also be extended to incorporate no-slip conditions, which may significantly influence particle interactions.
Finally, an important extension would be to account for large deformations, under which the contact network can evolve and new contacts can form between neighbouring particles. This would enable the model to describe changes in packing geometry and their effects on the macroscopic response.

\appendix

\section{Spectral decomposition of the Jacobians}
\label{sec:app A}
We report the expressions of the Jacobian $\bb A(\bb U)$ of Eq.~\eqref{eq:motion} and its eigenvectors. We have
\begin{equation}\label{eq:Jacobian A}
\bb A =
\begin{bmatrix}
 0 & 1 & 0 & 0 & 0  & 0 & 0\\
 \kH\sum_i\A_i-v_1^2 & 2 v_1 & 0  &  A_{2,4} & A_{2,5}  & A_{2,6}  &  A_{2,7}
 \\
 \kH\sum_i\frac{\A_i b_i^y}{b_i^x}-v_1 v_2 & v_2 & v_1  &   A_{3,4} & A_{3,5}  & A_{3,6}  &  A_{3,7}
 \\
 0 & 0 & 0 & v_1 & 0 & 0 & 0 \\
 0 & 0 & 0 & 0 & v_1 & 0 & 0 \\
 F_{11} v_2-F_{21} v_1 & F_{21} & -F_{11} & 0 &  0 & v_1 & 0 \\
 F_{12} v_2-F_{22} v_1 & F_{22} & -F_{12} & 0 &  0 & 0 & v_1
\end{bmatrix}
\end{equation}
where
\begin{align}
    A_{2,4} &= \kH\sum_i
          \frac{a_i^xb_i^x\xi_i^{n-1}}{|\bb b_i|^3}\Big(
          \big(|\bb b_i|^2 + (b_i^y)^2\big)\xi_i - (b_i^x)^2 n|\bb b_i|
          \Big)
     ,&
      A_{3,4} &= \kH\sum_i
          \frac{a_i^x b_i^y\xi_i^{n-1}}{|\bb b_i|^3} \big((b_i^y)^2\xi_i - (b_i^x)^2n|\bb b_i|\big)
      ,\notag\\
     A_{2,5} &= \kH\sum_i
           \frac{a_i^yb_i^x\xi_i^{n-1}}{|\bb b_i|^3}\Big(
           \big(|\bb b_i|^2+(b_i^y)^2\big) \xi_i -n(b_i^x)^2|\bb b_i|
           \Big)
      ,&
      A_{3,5} &= \kH\sum_i
          \frac{a_i^y b_i^y\xi_i^{n-1}}{|\bb b_i|^3} \big((b_i^y)^2\xi_i-n(b_i^x)^2|\bb b_i|\big)
     ,\notag
\end{align}
\begin{align}
      A_{2,6} &=-\kH\sum_i
          \frac{a_i^x b_i^y(b_i^x)^2\xi_i^{n-1}}{|\bb b_i|^3}(n|\bb b_i| +\xi_i)
      ,&
      A_{3,6} &= \kH\sum_i
          \frac{a_i^xb_i^x\xi_i^{n-1}}{|\bb b_i|^3} \big((b_i^x)^2 \xi_i - n(b_i^y)^2|\bb b_i|\big)
      ,\notag\\
      A_{2,7} &= -\kH\sum_i
          \frac{a_i^y b_i^y(b_i^x)^2\xi_i^{n-1}}{|\bb b_i|^3}(n|\bb b_i| + \xi_i)
      ,&
      A_{3,7} &= \kH\sum_i
          \frac{a_i^yb_i^x\xi_i^{n-1}}{|\bb b_i|^2} \big((b_i^x)^2 \xi_i - n(b_i^y)^2|\bb b_i|\big).
\end{align}

\noindent
The right-eigenvectors $\bb r_j$ of $\bb A$ corresponding to the eigenvalues $\lambda_i$ given in Eq.~\eqref{eq:eigenvalues 3 cont} are
\begin{align}
    \bb r_1&=[1,\;v_1,\;v_2,\;0
    ,\;
    \mathcal D/\mathcal C
    ,\;0,\;
    \mathcal E/\mathcal C
    ],
    \notag\\[-4pt]
    \bb r_2&=\sum_i \frac{a_i^xb_i^x\xi_i^{n-1}}{|\bb b_i|^7}      \Big[0,\;0,\;0,
    \;
    n(b_i^y)^2 |\bb b_i|-(b_i^x)^2\xi_i
    ,\;0,\;
     \big((b_i^y)^2+|\bb b_i|^2\big) \xi_i  - n(b_i^x)^2|\bb b_i|
    ,\;
    0\Big],
    \notag\\[-4pt]
    \bb r_3 &= \sum_i    \frac{a_i^x b_i^x\xi_i^{n-1}}{|\bb b_i|^3}    \Big[0,\;0,\;0,\;
     \frac{a_i^y b_i^y}{a_i^x b_i^x}\Big(n(b_i^x)^2|\bb b_i| - (b_i^y)^2 \xi_i\Big)
    ,\;
     n(b_i^x)^2|\bb b_i| - \big((b_i^y)^2+|\bb b_i|^2\big) \xi_i
    ,\;
    0,\;0\Big],
    \notag\\[-4pt]
    \bb r_j&=
    \begin{bmatrix}
        1  			,&
        \lambda_j	,&
        v_2  + (\lambda_j-v_1) \,\mathcal F/\mathcal G_j  ,&
        0		,&
        0		,&
        F_{21} - F_{11} \,\mathcal F/\mathcal G_j   ,&
        F_{22} - F_{12} \,\mathcal F/\mathcal G_j
    \end{bmatrix}^\top,
    \label{eq:r eigvects}
\end{align}
where $j\in\{4,5,6,7\}$ and 
\begin{subequations}
\begin{align}
    \mathcal C&={A_{2,7} A_{3,5} - A_{2,5} A_{3,7}},
    \qquad\quad
    \mathcal D = 
    A_{3,7}  \sum_i \A_i - A_{2,7} \sum_i \A_i \frac{b_i^y}{b_i^x},
    \\
    \mathcal E &= 
    -A_{3,5} \sum_i\A_i +  A_{2,5} \sum_i \A_i {b_i^y}{b_i^x},
    \\
	\mathcal F &= 
     \kH\sum_i\frac{b_i^x b_i^y}{|\bb b_i|^2}\big(\A_i+\B_i\big),
    \qquad
	\mathcal G_j = (\lambda_j-v_1)^2
    +\kH\sum_i\frac{1}{|\bb b_i|^2} \big((b_i^x)^2\B_i-(b_i^y)^2 \A_i\big),
\end{align}
\end{subequations}
When three directions of contact are active, $\A_i,\,\B_i>0$ and $\lambda_{4,5,6,7}$ are distinct; thus, the eigenvectors $r_{4,5,6,7}$ are linearly independent.
It is immediate to see that $\bb r_1$ is linearly independent with respect to $\bb r_2$ and $\bb r_3$.
Finally, $\bb r_2$ and $\bb r_3$ are dependent only if one of the two is the null vector or the fifth component of $\bb r_2$ and the fourth component of $\bb r_3$ are zero simultaneously. Numerical checks confirm that this does not happen in any of the simulations conducted in this paper.
Note that it may happen that $\mathcal F=\mathcal G_j=0$; in this case, $\bb r_j$ is obtained by defining $\mathcal F/\mathcal G_j=0$.

\subsection{System in quasi-conservative form}
\label{sec:app B}

We perform the spectral decomposition of the Jacobian matrices of system \eqref{eq:motion num}. Along the direction $x$, we obtain the same matrix as \eqref{eq:Jacobian A}, except a few terms in columns four and five: the terms in positions $(4,4)$ and $(5,5)$ are zeros, and the terms in $(6,4)$ and $(7,5)$ are $-v_2$. The right eigenvectors $\bb r_{4,5,6,7}$ provided by \eqref{eq:r eigvects} are null in position four and five, and thus they are also right eigenvectors of $\tilde{\bb A}$, along with the corresponding eigenvalues $\lambda_{4,5,6,7}$ provided by Eq.~\eqref{eq:eigenvalues 3 cont}.
\\
The remaining eigenvalues are
$\tilde\lambda_1 = v_1$ and $\tilde\lambda_{2,3} = 0$.
However, the corresponding eigenvectors cannot, in general, be chosen continuously with respect to the state $\bb U$. Indeed, 
a discontinuity has been found at $(v_1=0;\,v_2=0)$.
This ill-conditioned eigenvalue problem is particularly problematic in the present setting, since the solutions are computed close to the steady state.

\subsection{Split systems}
\label{sec:app C}

In this section we compute the eigenvalues and eigenvectors of the two systems obtained by splitting system~\eqref{eq:motion num}. Written in quasi-linear form, we obtain
\begin{equation}
    \begin{bmatrix}
        \bb U_1  \\ \bb U_2
    \end{bmatrix}_t +
    \begin{bmatrix}
        \bb A_{11}  & \bb A_{12}
        \\
        \bb A_{21}  & \bb A_{22}
    \end{bmatrix}
    \begin{bmatrix}
        \bb U_1  \\ \bb U_2
    \end{bmatrix}_x +
    \begin{bmatrix}
        \bb B_{11}  & \bb B_{12}
        \\
        \bb B_{21}  & \bb B_{22}
    \end{bmatrix}
    \begin{bmatrix}
        \bb U_1  \\ \bb U_2
    \end{bmatrix}_y
    =
    \begin{bmatrix}
        \bb F_1  \\ \bb F_2
    \end{bmatrix}_x
\end{equation}
where $\bb U_1 = [\rho, \rho v_1, \rho v_2]^\top$ and $\bb U_2 = [\rho F_{11}, \rho F_{12}, \rho F_{21}, \rho F_{22}]^\top$.
We split the problem and consider the spectral decomposition of $\bb A_{11}$ and $\bb A_{22}$, and of $\bb B_{11}$ and $\bb B_{22}$.
For $\bb A_{11}$ we have
\begin{subequations}
    \begin{align}
        (\tilde\lambda_1)_1 &= v_1,
        &(\tilde{\bb r}_1)_1 &= [0,\,0,\,1]^\top,
        \\[-8pt]
        (\tilde\lambda_1)_{2,3} &= v_1 \pm \sqrt{\sum\nolimits_i \A_i}\;,
        &(\tilde{\bb r}_1)_{2,3} &= \Big[1,\;(\tilde\lambda_1)_{2,3},\; \pm\frac{\sum_i \frac{b_i^y}{b_i^x}\A_i}{\sum_i \A_i}\Big]^\top.
    \end{align}
\end{subequations}
Analogue results are obtained for $\bb B_{11}$.
These are used to implement the $f$-splitting scheme.
\\
The spectral decomposition of $\bb A_{22}$ and $\bb B_{22}$ is not necessary for the numerics because the Rusanov scheme inside the MUSCL--Hankock only needs an estimation of the maximum wave speed. The eigenvalues of the full system provided by Eq.~\eqref{eq:eigenvalues 3 cont} are used.

\section{Numerical discretisation}

By looking for a numerical solution, particular care has been spent to approximate the source term due to gravity. Due to its general application, we rely on the method proposed by Bale and LeVeque \cite{bale2003wave}, called flux-splitting.
This scheme decomposes the jump of the numerical flux across each cell interface into a set of propagating waves, or $f$-waves, whose amplitudes are determined directly from the flux difference. In contrast to a conventional characteristic decomposition of the state jump \cite{leveque1997wave}, the source term can be incorporated directly into the flux difference before the wave decomposition. The resulting fluctuations therefore account simultaneously for the propagation of the physical flux and for the local action of the source term. This provides a natural mechanism to capture the balance between flux gradients and source terms without requiring an explicit expression for the equilibrium solution.
As a drawback, this scheme requires to perform a spectral decomposition of the flux Jacobians in each cell at every time step.
The numerical search of eigenvectors resulted to be numerically demanding and, for certain states, ill posed, particularly for values of velocity close to zero. See Appendix~\ref{sec:app B} for more details.

To address this complication, the system has been split using Goudunov splitting \cite{leveque1992numerical,toro2013riemann} into two main hyperbolic steps, one solved after the other for each time step. The first step propagates the mass \eqref{eq:motion num-a} and the momentum \eqref{eq:motion num-b} conservation. By assuming the value of $\rho\bb F$ as a given field, constant during the time step, these two equations are solved using the $f$-splitting. This allows to account for the source with a precision closed to second order. The spectral decomposition needed for this step is described in Appendix~\ref{sec:app C}.
\\
The second step solves Eq.~\eqref{eq:motion num-c} to evolve the deformation; during this time step, the states $\rho$ and $\rho\bb v$ are considered as time-constant fields.
Eq.~\eqref{eq:motion num-c} is discretized using the second-order MUSCL--Hancock scheme with the MINMOD flux limiter \cite{toro2013riemann}, while numerical fluxes are computed using the Rusanov approximate Riemann solver. For this second step, a dimensional splitting approach is adopted, separating the $x$- and $y$-fluxes as well as the source term. The latter is integrated with a fourth-order Runge--Kutta scheme, where the divergence $\nablax\cdot(\rho\bb F^\top)$ is computed with a centred finite difference. Here, a precise resolution of the source term is not fundamental because it is very small and it vanishes exactly at the static equilibrium.
Although formally these multiple splittings converge to first order, an empirical convergence very close to second order has been found.
\\
Explicit time integration is exploited, with the time step selected according to a CFL condition such that \cbox{$\mathrm{CFL}\le0.6$}, exploiting the availability of the exact characteristic wave speeds of the whole system~\eqref{eq:motion num}. These are very similar to those of system~\eqref{eq:motion} as detailed in Appendix~\ref{sec:app B}.
\\
The illustrative example on the spurious oscillations obtained without the well-balanced scheme has been obtained using the scheme described above, but the source term due to gravity and friction is solved via Runge-Kutta using a further splitting.

Boundary conditions are imposed using the ghost-cell approach \cite{leveque1992numerical}.
For instance, the state variables at the ghost-cells on $\Gamma_\mathrm{L}$ are set so that symmetry is implied across the boundary:
\begin{align}\label{eq:BCs num}
    &\rho_\mathrm{GL}=\rho\big|_{\Gamma_\mathrm{L}}\,, \qquad\quad\,\,
    (\rho v_1)_\mathrm{GL}=\rho (2v_\mathrm{L}- v_1)\big|_{\Gamma_\mathrm{L}}\,, &&
    (\rho v_2)_\mathrm{GL}=\rho v_2\big|_{\Gamma_\mathrm{L}}\,,
    \\\notag
    &(\rho F_{11})_\mathrm{GL}=\rho F_{11}\big|_{\Gamma_\mathrm{L}}\,, \quad
    (\rho F_{12})_\mathrm{GL}=-\rho   
    F_{12}\big|_{\Gamma_\mathrm{L}}\,, &&
    (\rho F_{21})_\mathrm{GL}=-\rho F_{21}\big|_{\Gamma_\mathrm{L}}\,, &&
    (\rho F_{22})_\mathrm{GL}=\rho F_{22}\big|_{\Gamma_\mathrm{L}}\,,
\end{align}
where the subscript $\mathrm{GL}$ indicates values of the ghost cell on the left boundary. 
Note that the deformation gradient in the ghost-cells $\bb F_\mathrm{GL}$ is changed so that a symmetry condition holds on the boundaries. Similar boundary conditions are applied to $\Gamma_\mathrm{R}$, $\Gamma_\mathrm{B}$, and $\Gamma_\mathrm{T}$.
Two ghost-cells are used on each side of the domain to approximate the boundary conditions at the second order. Thus, the condition \eqref{eq:BCs num} is applied symmetrically with respect to the boundary: between the last physical cell and the inner ghost-cell, and between the second last physical cell and the outer ghost cell.
The $f$-wave method solves the Riemann problems at the cell level by considering the contribution of the source as concentrated at the inter-cell boundaries. Thus, it is important to symmetrise also the source terms into the ghost cells. In practice, the component of $\rho\bb g-\gamma\rho\bb v$ normal to the boundary changes sign in the corresponding ghost cell.

Lastly, we remark that the conservation of mass provided by Eq.~\eqref{eq:motion num-a} is, in principle, redundant, since the density can be recovered from the relation $\rho\det\bb F=\rho_{0,\mathrm{hex}}$. However, at the discrete level, this procedure does not guarantee exact mass conservation. We therefore compute the volume changes directly from Eq.~\eqref{eq:motion num-a}, while evaluating the remaining components of the deformation using $\bb F$ from Eq.~\eqref{eq:motion num-c}. In particular, when evaluating the internal energy and stress tensor, we rescale the deformation gradient $\bb F$ such that its determinant is consistent with the volume implied by the density. Specifically, we apply the rescaling
$\bb F \rightarrow \frac{\rho_{0,\mathrm{hex}}}{\rho\det\bb F}\bb F$. A more detailed discussion of a similar approach can be found in Favrie and Lombard~\cite{favrie2023hyperbolic}.

\paragraph{Code availability.} The \Matlab{} code used to implement the numerical scheme and generate the results presented in this article is available at \url{https://github.com/SebaComi/Waves_granular_material}.

\ack{
S.C.\ received support from the French government under the France 2030 investment plan, as part of the Initiative d'Excellence d'Aix-Marseille Université - A*MIDEX(AMX-19-IET-010).
R.C.\ acknowledges the support of the French Agence Nationale de la Recherche (ANR), under grant number ANR-24-CE46-4528 (OREA project).
The authors thank Walter Boscheri and Nicolas Favrie for their valuable discussions and suggestions.
}

\printbibliography

\end{document}